\documentclass[aps,prd,preprint,floatfix,superscriptaddress,nofootinbib,longbibliography,a4paper]{revtex4-1}
\pdfoutput=1
\usepackage{color}
\usepackage{graphicx}
\usepackage{textcomp}
\usepackage{subfigure}
\usepackage{feynmp}
\usepackage{amsmath,amssymb,array}
\usepackage{enumerate}
\usepackage{hhline}
\usepackage{hyperref}
\hypersetup{
    citecolor= red,
    colorlinks=true,
    linkcolor=blue,
    filecolor=magenta,      
    urlcolor=blue,}

\newcommand{\beqa}{\begin{eqnarray}}
\newcommand{\eeqa}{\end{eqnarray}}
\newcommand{\be}{\begin{equation}}
\newcommand{\ee}{\end{equation}}
\newcommand{\ba}{\begin{array}} 
\newcommand{\ea}{\end{array}}
\newcommand{\bds}{$B^0_{d/s}$-$\bar{B}^0_{d/s}$\,}

\newcommand{\kk}{$K^0$-$\bar{K}^0$\,}
\newcommand{\dd}{$D^0$-$\bar{D}^0$\,}

\begin{document} 
\vspace*{0.5cm}
\title{Spectrum of colour sextet scalars in realistic SO(10) GUT}
\bigskip
\author{Ketan M. Patel}
\email{kmpatel@prl.res.in}
\affiliation{Theoretical Physics Division, Physical Research Laboratory, Navarangpura, Ahmedabad-380009, India}
\author{Saurabh K. Shukla}
\email{saurabhks@prl.res.in}
\affiliation{Theoretical Physics Division, Physical Research Laboratory, Navarangpura, Ahmedabad-380009, India}
\affiliation{Indian Institute of Technology Gandhinagar, Palaj-382355, India \vspace*{1cm}}

\begin{abstract}
Incorporation of the standard model Yukawa interactions in a grand unified theory (GUT) often predicts varieties of new scalars that couple to the fermions and lead to some novel observational effects. We assess such a possibility for the colour sextet diquark scalars within the realistic renormalizable models based on $SO(10)$ GUT.  The spectrum consists of five sextets: $\Sigma \sim (6,1,-\frac{2}{3})$, $S \sim (6,1,\frac{1}{3})$, $\overline{S}\sim(\overline{6},1,-\frac{1}{3})$, ${\cal S}\sim(6,1,\frac{4}{3})$ and $\mathbb{S}\sim(\overline{6},3,-\frac{1}{3})$. Computing explicitly their couplings with the quarks, we evaluate their contributions to the neutral meson-antimeson mixing and baryon number-violating processes like neutron-antineutron oscillation. The latter arises because of a $B-L$ violating trilinear coupling between the sextets which also contributes to some of the quartic couplings and perturbativity of the same leads to strong limits on the sextet masses. Using the values of the $B-L$ breaking scale and Yukawa couplings permitted in the realistic models, we derive constraints on the masses of these scalars. It is found that $\Sigma$ along with any of the remaining sextets cannot be lighter than the $B-L$ breaking scale, simultaneously. In the realm of realistic models, this implies no observable $n$-$\bar{n}$ oscillation in near future experiments. We also point out a possibility in which sub-GUT scale $\Sigma$ and a pair of $S$, allowed by the other constraints, can viably produce the observed baryon asymmetry of the universe.
\end{abstract}

\maketitle

\section{Introduction}
\label{sec:intro}
Grand Unified Theories (GUTs), which provide complete unification of the Standard Model (SM) gauge bosons and full or partial unification of the quarks and leptons, typically predict an enlarged spectrum for spin-0 particles \cite{Fritzsch:1974nn,Georgi:1974sy,Gell-Mann:1979vob}. This is in particular the case for the renormalizable versions of $SO(10)$ models constructed on the four-dimensional spacetime which provide a unique platform for constructing an explicit, predictive and realistic model of the grand unification \cite{Dimopoulos:1991yz,Babu:1992ia,Clark:1982ai,Aulakh:1982sw,Aulakh:2003kg}. Several scalar fields with varieties of colour and electroweak charges are predicted as partners of the electroweak Higgs doublets in these models. Since only the latter are essentially required in the low energy theory to break the electroweak symmetry, the rests are often assumed as heavy as the GUT scale invoking the so-called minimal survival hypothesis \cite{delAguila:1980qag}. Nevertheless, if some of these scalars remain lighter than the GUT scale then they can give rise to some phenomenologically interesting effects because of their non-trivial SM charges and direct couplings with quarks and leptons as predicted by the underlying GUT model. Such effects include flavour anomalies \cite{Bordone:2016gaq,Belanger:2021smw,Perez:2021ddi,Sahoo:2021vug,Aydemir:2022lrq}, distinct signatures for nucleon decays \cite{Patel:2022wya}, neutron-antineutron oscillation \cite{Kuzmin:1970nx,Ma:1998pi,Babu:2012vb,Babu:2012vc,Fridell:2021gag}, baryogenesis \cite{Babu:2012vb,Babu:2012vc,Babu:2012iv,Enomoto:2011py,Gu:2011ff,Gu:2017cgp,Hati:2018cqp}, precise unification of the SM gauge couplings \cite{FileviezPerez:2008afb,Dorsner:2009mq,Patel:2011eh,Babu:2012iv,Babu:2012vb} and some anomalous events in the direct search experiments \cite{Dorsner:2009mq,Patel:2011eh,Dorsner:2016wpm}.

A complete classification of the scalars that may arise from the most general Yukawa sector of the renormalizable $SO(10)$ models is given in our previous paper \cite{Patel:2022wya}. Among the various scalars, the colour triplet and sextet fields are of particular phenomenological interest as they all carry non-zero $B-L$, where $B$ ($L$) denotes Baryon (Lepton) number. Because of this, they give rise to processes that violate $B$ and/or $L$ which otherwise are good accidental global symmetries of the SM at the perturbative level. Among these, the colour triplets are known to induce nucleon decay and they have been comprehensively studied in \cite{Patel:2022wya}. Computing explicitly their couplings with the quarks and leptons in the realistic $SO(10)$ GUTs, we derived bounds on their masses arising from various $B-L$ conserving and violating modes of proton and neutron decays.  In this paper, we focus on the colour sextet scalars with a similar intention to derive the constraints on their spectrum from various phenomenological considerations.

Unlike the colour triplet scalars, the sextets do not induce nucleon decay by themselves. However, they can give rise to neutral baryon-antibaryon oscillations if there exists $B-L$ violating interaction between the relevant sextets \cite{Ma:1998pi,Arnold:2012sd,FileviezPerez:2015mlm}. The latter is inherent in the renormalizable $SO(10)$ models. One, therefore, expects constraints on the masses of the sextet scalars from neutron-antineutron oscillation experiments \cite{Baldo-Ceolin:1994hzw,Super-Kamiokande:2020bov,Addazi:2020nlz,DUNE:2020ypp}. Light colour sextets can also be constrained from the $B$ conserving but flavour-violating mixings between mesons and antimesons \cite{Ma:1998pi,Babu:2013yca,Giudice:2011ak,Mohapatra:2007af}. Both these constraints primarily depend on (i) the Yukawa couplings of the quarks with the underlying sextet scalars and (ii) the $B-L$ breaking scale. Unlike in the typical bottom-up approaches, both (i) and (ii) are more or less determined from the low energy spectrum of the quarks and leptons in the realistic renormalizable $SO(10)$ models. Thus, one obtains more robust and unambiguous bounds on the spectrum of the coloured sextet scalars in the top-down approach that we present in this work. Utilizing the spectrum of the sextet scalars allowed within the renormalizable $SO(10)$ models, we also point out a new and self-sufficient possibility of generating the observed baryon asymmetry of the universe.

The rest of the paper is organized as the following. In the next section, we derive the spectrum and couplings of colour sextet scalars in renormalizable $SO(10)$ models. Various phenomenological implications of these scalars are derived which include flavour violation in section \ref{sec:qfv}, neutron-antineutron oscillations in section \ref{sec:nnbar}, perturbativity of the effective quartic couplings in section \ref{sec:quartic} and baryogenesis in section \ref{sec:baryo}. Constraints from all these observables are analysed in section \ref{sec:results} and the study is concluded in section \ref{sec:concl}.

\section{Colour sextet scalars and their couplings}
\label{sec:couplings}
The Yukawa sector of renormalizable $SO(10)$ GUTs comprises scalars in ${\bf 10}$, $\overline{\bf 126}$ and ${\bf 120}$ dimensional irreducible representations of the gauge group. Various submultiplets residing in  these GUT multiplets, along with their SM and $B-L$ charges and multiplicities, are listed in our previous paper \cite{Patel:2022wya}. For convenience, we reproduce the information relevant to $SU(3)_C$ sextet fields in Table \ref{tab:scalars}. These scalars arise only from $\overline{\bf 126}_H$ and ${\bf 120}_H$.
\begin{table}[t]
\begin{center}
\begin{tabular}{cccccc} 
\hline
\hline
~~SM charges~~&~~Notation~~&~~$B-L$~~&~~$\overline{\bf 126}_H$~~&~~${\bf 120}_H$~~\\
 \hline
$\left(6,1,\frac{1}{3}\right)$ & $S^{\alpha}_{\beta\gamma}$ & $\frac{2}{3}$  &1&1\\
$\left(\overline{6},1,-\frac{1}{3}\right)$&$\overline{S}{^{\beta\gamma}_{\alpha}}$ &$-\frac{2}{3}$  &0 &1 \\
$\left(6,1,-\frac{2}{3}\right)$ & $ \Sigma^{\alpha\beta}$ & $\frac{2}{3}$  &1 &0 \\
$\left(6,1,\frac{4}{3}\right)$ & ${\cal S}^{\alpha}_{\beta\gamma}$ & $\frac{2}{3}$  & 1 & 0 \\
$\left(\overline{6},3,-\frac{1}{3}\right)$ &$\mathbb{S}^{\alpha\beta a}_{\gamma b}$ & $-\frac{2}{3}$ &1 &0\\
\hline
\end{tabular}
\end{center}
\caption{Types of coloured sextet fields, their charges under  the SM gauge group ($SU(3)_C$, $SU(2)_L$, $U(1)_Y$), $B-L$ and  multiplicities in $\overline{\bf 126}_H$ and ${\bf 120}_H$ dimensional scalars  of $SO(10)$.}
\label{tab:scalars}
\end{table}

The couplings of these coloured sextet fields with the SM quarks can be straightforwardly computed using the method discussed and the decompositions given in \cite{Patel:2022wya}. For the sextet fields residing in $\overline{\bf 126}_H$, we find 
\beqa \label{sextet_126}
-{\cal L}^{\overline{\bf 126}}_Y &=& F_{AB}\,{\bf 16}^T_A\,C^{-1}\,{\bf 16}_B\,\overline{\bf 126}_H\,+\,{\rm h.c.}\, \nonumber \\
& \supset & -\frac{i}{\sqrt{15}}\,F_{AB}\,\Big(2\, d^{C T}_{\alpha A}\,C^{-1}\,d^C_{\beta B}\, \Sigma^{\alpha \beta}\,-\, \sqrt{2}\, \epsilon^{\alpha \beta \gamma}\, u^{C T}_{\gamma A}\,C^{-1}\,d^C_{\sigma B}\, {S}^\sigma_{\alpha \beta} \Big. \nonumber \\
& & \Big. + \epsilon^{\alpha \beta \gamma}\, u^{C T}_{\sigma A}\,C^{-1}\,u^C_{\gamma B}\,{\cal S}^\sigma_{\alpha \beta}\,+\,\sqrt{2}\, \epsilon_{\alpha \beta \gamma}\, \epsilon_{ab}\, q^{a \alpha T}_{A}\,C^{-1}\,q^{\sigma c}_B\,\mathbb{S}^{\beta \gamma b}_{\sigma c}\Big)\,+ \, {\rm h.c.}\,.\eeqa
In the above, we continue following the notations used by us in the previous work \cite{Patel:2022wya} in which the $\alpha,\beta,...$ ($a,b,...$) letters denote $SU(3)_C$ ($SU(2)_L$) indices while $A, B,...$ represent three flavours of quarks. Different numerical factors in front of each term in Eq. (\ref{sextet_126}) arise from the Clebsch-Gordan decomposition and canonical normalization of the quark and scalar fields \cite{Patel:2022wya}.

Analogously for the Yukawa interaction with ${\bf 120}_H$, we obtain
\beqa \label{sextet_120}
-{\cal L}^{\bf 120}_Y &=& G_{AB}\,{\bf 16}^T_A\,C^{-1}\,{\bf 16}_B\,{\bf 120}_H\,+\,{\rm h.c.}\, \nonumber \\
& \supset & -\frac{2i}{\sqrt{3}}\,G_{AB}\,\Big(\epsilon^{\alpha \beta \gamma}\, u^{C T}_{\gamma A}\,C^{-1}\,d^C_{\sigma B}\, \tilde{S}^\sigma_{\alpha \beta} - \epsilon_{\alpha \beta \gamma}\, u^{\gamma T}_A\,C^{-1}\,d^\sigma_B\, \overline{S}^{\alpha \beta}_\sigma \Big)\,+ \, {\rm h.c.}\,.\eeqa
Here, $F$ and $G$ are symmetric and antisymmetric matrices in the flavour space, respectively. $\tilde{S}$ denotes the color sextet with $Y=1/3$ residing in ${\bf 120}_H$ and it is distinguished from $S$ belonging to $\overline{\bf 126}_H$ which has the same quantum numbers. It is noted that $\overline{S}$ and $\mathbb{S}$ couple to the left-chiral quark fields while the remaining colour sextets have interaction vertices with only the right-chiral quarks. All the interactions in Eqs. (\ref{sextet_126},\ref{sextet_120}) conserve $B-L$.

In the models with both $\overline{\bf 126}_H$ and ${\bf 120}_H$ present, the fields $S$ and $\tilde{S}$ can mix with each other through gauge invariant terms like ${\bf 120}_{H}\overline{{\bf 126}}^{\dagger}_{H}{\bf 45}_{H}$ or ${\bf 120}_{H}\overline{{\bf 126}}^{\dagger}_{H}{\bf 210}_{H}$. The physical states are then given by linear combinations of $S$ and $\tilde{S}$. For simplicity, we assume that such linear combinations are parametrized by real parameters and define
\be \label{S12}
S_1 = c_\theta\, S\,+\,s_\theta\,\tilde{S}\,,~~ S_2 = -s_\theta\, S\,+\,c_\theta\,\tilde{S}\,.\ee
Here, $c_\theta = \cos \theta$ and $s_\theta = \sin \theta$. The linear combination $S_1$ is to be identified with the lighter mass eigenstate, i.e. $M_{S_1} < M_{S_2}$.

Replacing $S$ and $\tilde{S}$ with the physical states $S_{1,2}$ using Eq. (\ref{S12}) and converting the quarks fields into their physical basis using $f \to U_f\, f$, the Yukawa couplings between the various sextet fields, $\Phi=\Sigma, S_i,{\cal S},\mathbb{S},\overline{S}$, and quarks can be rewritten as
\beqa \label{sextet_all}
-{\cal L}_\Phi & = & Y^\Sigma_{AB}\,d^{C T}_{\alpha A}\,C^{-1}\,d^C_{\beta B}\, \Sigma^{\alpha \beta}\,+\, Y^{S_i}_{AB}\,\epsilon^{\alpha \beta \gamma}\, u^{C T}_{\gamma A}\,C^{-1}\,d^C_{\sigma B}\, {S_i}^\sigma_{\alpha \beta}\, \nonumber \\
& + & Y^{\cal S}_{AB}\,\epsilon^{\alpha \beta \gamma}\, u^{C T}_{\sigma A}\,C^{-1}\,u^C_{\gamma B}\,{\cal S}^\sigma_{\alpha \beta}\,+\,Y^{\mathbb{S}}_{AB}\, \epsilon_{\alpha \beta \gamma}\, \epsilon_{ab}\, q^{a \alpha T}_{A}\,C^{-1}\,q^{\sigma c}_B\,\mathbb{S}^{\beta \gamma b}_{\sigma c} \nonumber \\
& + & Y^{\overline{S}}_{AB}\,\epsilon_{\alpha \beta \gamma}\, u^{\gamma T}_A\,C^{-1}\,d^\sigma_B\, \overline{S}^{\alpha \beta}_\sigma\,+\,{\rm h.c.}\,,\eeqa
where $i=1,2$. The $3 \times 3$ matrices $Y^\Phi$ obtained from Eqs. (\ref{sextet_126},\ref{sextet_120}) as
\beqa \label{YPhi}
Y^\Sigma &=& -\frac{2 i}{\sqrt{15}}\,U_{d^C}^T\, F\, U_{d^C}\,,~Y^{S_1} = \frac{\sqrt{2}i}{\sqrt{3}}\,\left(\frac{1}{\sqrt{5}} c_\theta\, U_{u^C}^T\,F\,U_{d^C}\, -\, \sqrt{2}s_\theta\, U_{u^C}^T\,G\,U_{d^C}\right)\,, \nonumber \\
Y^{S_2} &=& -\frac{\sqrt{2}i}{\sqrt{3}}\,\left(\frac{1}{\sqrt{5}} s_\theta\, U_{u^C}^T\,F\,U_{d^C}\, +\, \sqrt{2}c_\theta\, U_{u^C}^T\,G\,U_{d^C}\right)\,,~Y^{\cal S}  =  -\frac{i}{\sqrt{15}}\,U_{u^C}^T\,F\,U_{u^C}\,, \nonumber \\
Y^{\mathbb{S}} & = & -\frac{\sqrt{2} i}{\sqrt{15}}\,U_{q}^T\,F\,U_{q^\prime}\,,~Y^{\overline{S}}  = \frac{2 i}{\sqrt{3}}\,U_u^T\,G\,U_d \,.\eeqa
Here $q,q^\prime = u,d$. The matrices $Y^\Sigma$, $Y^{\cal S}$ are symmetric in the flavour space while $Y^{\mathbb{S}}$ is symmetric when $q=q^\prime$. Note that the matrices $U_u$ and $U_d$ determine the quark mixing matrix, $U_u^\dagger U_d \equiv V_{\rm CKM}$. Therefore, $U_u \sim U_d$ serves as a good approximation and $Y^{\mathbb{S}}$ can be considered symmetric at the leading order.

The advantage of deriving the expressions in Eq. (\ref{YPhi}) is that all the $Y^\Phi$ can be explicitly computed in the realistic $SO(10)$ models in which the fundamental couplings $F$, $G$ and the diagonalizing matrices $U_f$ are determined from the fermion mass fits. We now use the above couplings to determine various phenomenologically relevant processes involving sextet scalars in the subsequent sections.

\section{Quark flavour violation}
\label{sec:qfv}
The Yukawa sector of viable $SO(10)$ models must consist of more than one GUT scalar. This implies that $Y^\Phi$ are not diagonal matrices and, therefore, the sextet scalars can lead to a new source of flavour violation in the quark sector. The strongest constraints on this type of new physics come from the $|\Delta F|=2$ processes involving neutral meson-antimeson oscillations. Following the effective theory approach, we estimate the sextets-induced contributions to \kk, \bds and \dd mixing at tree and 1-loop levels.

Integrating out various sextets from Eq. (\ref{sextet_all}) and parametrizing the effective Lagrangian as
\beqa \label{L_F2}
{\cal{L}}^{\Delta F=2}_{\rm eff} &=& \sum _{q=d,u} \left(c_q\, {\cal{O}}_q\;+\; \tilde{c}_q\, {\cal{\tilde{O}}}_q \right)\,+\, \rm{h.c.}\,,\eeqa
we find the following independent operators
\beqa \label{O_F2}
{\cal{\tilde{O}}}_{d}\;&=&\;\Big(\overline{d{_{R}^{\alpha}}}_{A} \gamma^{\mu}\,d{_{R\,C}^{\alpha}}\Big)\,\left(\overline{d{_{R}^{\beta}}}_{B}\,\gamma_{\mu}\,d{_{R\,D}^{\beta}}\right)\,, \nonumber \\
{\cal{\tilde{O}}}_{u}\;&=&\;\Big(\overline{u{_{R}^{\alpha}}}_{A} \gamma^{\mu}\,u{_{R\,C}^{\alpha}}\Big)\,\left(\overline{u{_{R}^{\beta}}}_{B}\,\gamma_{\mu}\,u{_{R\,D}^{\beta}}\right)\,, \nonumber \\
{\cal O}_{d}\;&=&\;\Big(\overline{d{_{L}^{\alpha}}}_{A} \gamma^{\mu}\,d{_{L\,C}^{\alpha}}\Big)\,\left(\overline{d{_{L}^{\beta}}}_{B}\,\gamma_{\mu}\,d{_{L\,D}^{\beta}}\right)\,, \nonumber \\
{\cal O}_{u}\;&=&\;\Big(\overline{u{_{L}^{\alpha}}}_{A} \gamma^{\mu}\,u{_{L\,C}^{\alpha}}\Big)\,\left(\overline{u{_{L}^{\beta}}}_{B}\,\gamma_{\mu}\,u{_{L\,D}^{\beta}}\right)\,.\eeqa
Here, we use $q = q_L$ and $q^C = C q_R^*$ for $q=u,d$ to obtain the above operators in the usual left- and right-chiral notations. ${\cal O}_{u,d}$ and ${\cal \tilde{O}}_{u,d}$ are related by $L \leftrightarrow R$. Operators ${\cal O}_u$ and ${\cal O}_d$ induce flavour-changing neutral meson-antimeson oscillations in the up-type and down-type quark sectors, respectively.

The coefficient of the operator ${\cal O}_d$ is obtained as 
\beqa \label{cd1}
\tilde{c}_d [A,B,C,D] & = & \frac{1}{M_{\Sigma}^2}\left[\frac{1}{2} Y^\Sigma_{AB} (Y^{\Sigma}_{CD})^* -\frac{6}{64\pi^2}\left[\left(Y_2^\Sigma\right)_{AC}\left(Y_2^\Sigma\right)_{BD} + \left(Y_2^\Sigma\right)_{AD}\left(Y_2^\Sigma\right)_{BC} \right]\right]\nonumber\\
& - & \frac{1}{M^2_{S_1}}\,\frac{6}{16\pi^2} \left[\left(\bar{Y}_2^{S_1}\right)_{AC}\left(\bar{Y}_2^{S_1}\right)_{BD} + \left(\bar{Y}_2^{S_1}\right)_{AD}\left(\bar{Y}_2^{S_1}\right)_{BC}\right]\,, \eeqa
where $Y_2^\Phi = Y^\Phi Y^{\Phi \dagger}$ and $\bar{Y}_2^\Phi = Y^{\Phi \dagger} Y^\Phi$ are hermitian matrices. The first term in the above expression denotes the tree-level contribution mediated by $\Sigma$ while the second and third terms are contributions generated at 1-loop by the scalars $\Sigma$ and $S_1$, respectively. Note that we haven't included the contribution from $S_2$, which is of a similar kind as $S_1$, as it is generically expected to be suppressed since $M_{S_2} > M_{S_1}$.

Analogously, we find the following coefficients of the remaining operators.
\beqa \label{cu1}
\tilde{c}_u [A,B,C,D] & = & \frac{1}{M_{\mathcal{S}}^2}\left[Y^{\cal S}_{AB} (Y^{\cal S}_{CD})^* - \frac{6}{16\pi^2} \left[\left(Y_2^{\cal S}\right)_{AC}\left(Y_2^{\cal S}\right)_{BD} + \left(Y_2^{\cal S}\right)_{AD}\left(Y_2^{\cal S} \right)_{BC}\right]\right] \nonumber\\
& - & \frac{1}{M^2_{S_1}}\,\frac{6}{16\pi^2} \left[\left(Y_2^{S_1}\right)_{AC}\left(Y_2^{S_1}\right)_{BD} + \left(Y_2^{S_1}\right)_{AD}\left(Y_2^{S_1}\right)_{BC}\right]\,,\eeqa
\beqa \label{cd2}
c_d [A,B,C,D] &=& \frac{1}{M^2_{\mathbb{S}}}\,\left[Y^{\mathbb{S}}_{AB} (Y^{\mathbb{S}}_{CD})^* - \frac{30}{16\pi^2}\left[\left(Y_2^\mathbb{S}\right)_{AC}\left(Y_2^\mathbb{S}\right)_{BD} + \left(Y_2^\mathbb{S}\right)_{AD}\left(Y_2^\mathbb{S}\right)_{BC}\right]\right]\nonumber\\
& - & \frac{1}{M^2_{\overline{S}}}\frac{6}{16\pi^2}\,\left[\left(\bar{Y}_2^{\overline{S}}\right)_{AC}\left(\bar{Y}_2^{\overline{S}}\right)_{BD} + \left(\bar{Y}_2^{\overline{S}}\right)_{AD}\left(\bar{Y}_2^{\overline{S}}\right)_{BC}\right]\,, \eeqa 
and 
\beqa \label{cu2}
c_u [A,B,C,D]&=& \frac{1}{M^2_{\mathbb{S}}}\,\left[Y^{\mathbb{S}}_{AB} (Y^{\mathbb{S}}_{CD})^* - \frac{30}{16\pi^2}\left[\left(Y_2^\mathbb{S}\right)_{AC}\left(Y_2^\mathbb{S}\right)_{BD} + \left(Y_2^\mathbb{S}\right)_{AD}\left(Y_2^\mathbb{S}\right)_{BC}\right]\right]\nonumber\\
& - & \frac{1}{M^2_{\overline{S}}}\frac{6}{16\pi^2}\,\left[\left(Y_2^{\overline{S}}\right)_{AC}\left(Y_2^{\overline{S}}\right)_{BD} + \left(Y_2^{\overline{S}}\right)_{AD}\left(Y_2^{\overline{S}}\right)_{BC}\right]\,.\eeqa
It can be seen from the above expressions of $\tilde{c}_{u,d}$ and $c_{u,d}$ that only the sextet $\Sigma$ (${\cal S}$) gives rise to flavour violations in the down-type (up-type) quark sectors while  $\mathbb{S}$ contributes in both the sectors at tree level. Contributions of $S_1$ and $\overline{S}$ to the flavour violation arise only at the loop level. Nevertheless, the latter can be of a similar order  as that of the tree level depending on the hierarchical structure of the Yukawa couplings.

For a quantitative estimation of the constraint on the masses and couplings of sextet scalars, relevant $c_{u,d}$ and $\tilde{c}_{u,d}$ need to be evolved from the mass scale of the integrated-out  scalar, i.e. $\mu = M_\Phi$ where Eqs. (\ref{cd1} - \ref{cu2}) hold, down to the scale at which meson-antimeson mixing is experimentally determined. For \kk oscillation, the relevant scale is $\mu=2$ GeV and the RGE evolved effective coefficient for $M_\Phi > m_t$ is given by \cite{Ciuchini:1998ix}
\be \label{CK}
C^1_K = \left(0.82 - 0.016\, \frac{\alpha_s(M_\Phi)}{\alpha_s(m_t)}\right) \left(\frac{\alpha_s(M_\Phi)}{\alpha_s(m_t)} \right)^{0.29}\, c_d[1,1,2,2](M_\Phi)\,.\ee
The same equation is obtained for $\tilde{C}^1_K$ by replacing $c_d$ with $\tilde{c}_d$ in the above expression. Note that $C^1_K$ and $\tilde{C}^1_K$ are coefficients of effective operators, $Q_1$ and $\tilde{Q}_1$, given in \cite{Ciuchini:1998ix} which are identical to our ${\cal O}_d$ and ${\cal \tilde{O}}_d$, respectively, with $A=B=1$ and $C=D=2$.  These operators do not mix with the other operators through RGE evolution and, therefore, one obtains a relatively simple expression, Eq. (\ref{CK}).

Similar arguments also follow for \bds mixing. In this case, one obtains the relevant Wilson coefficients as
\beqa \label{CB}
C^1_{B_d} &=& \left(0.865 - 0.017\, \frac{\alpha_s(M_\Phi)}{\alpha_s(m_t)}\right) \left(\frac{\alpha_s(M_\Phi)}{\alpha_s(m_t)} \right)^{0.29}\, c_d[1,1,3,3](M_\Phi)\,, \nonumber \\
C^1_{B_s} &=& \left(0.865 - 0.017\, \frac{\alpha_s(M_\Phi)}{\alpha_s(m_t)}\right) \left(\frac{\alpha_s(M_\Phi)}{\alpha_s(m_t)} \right)^{0.29}\, c_d[2,2,3,3](M_\Phi)\,, \eeqa
at  $\mu=m_b$ \cite{Becirevic:2001jj}. In the case of the charm mixing governed by \dd oscillations, one finds \cite{UTfit:2007eik}
\be \label{CD}
C^1_D = \left(0.837 - 0.016\, \frac{\alpha_s(M_\Phi)}{\alpha_s(m_t)}\right) \left(\frac{\alpha_s(M_\Phi)}{\alpha_s(m_t)} \right)^{0.29}\, c_u[1,1,2,2](M_\Phi)\,,\ee
at $\mu=2.8$ GeV. Various coefficients then can be compared with the present limits obtained from a fit to the experimental data by UTFit collaboration \cite{UTfit:2007eik}. The present limits are 
\be \label{const_C}
\left| C^1_K \right| < 9.6 \times 10^{-13},\,\left| C^1_{B_d} \right| < 2.3 \times 10^{-11},\,\left| C^1_{B_s} \right| < 1.1 \times 10^{-9},\,
\left| C^1_D \right| < 7.2 \times 10^{-13}\,.\ee
The same upper bounds are also applicable on the corresponding $\tilde{C}^1$.

\section{Neutron-Antineutron oscillations}
\label{sec:nnbar}
The colour sextet fields can also induce a transition between the neutral baryons and their antiparticles. Unlike the flavour transitions discussed in the previous section, this requires a source of  baryon number violation. In renormalizable $SO(10)$, the latter naturally arises from gauge invariant quartic couplings between three sextets and an SM singlet residing in $\overline{\bf 126}_H$, namely $\sigma$, which carries $B-L = -2$ (see Table I in \cite{Patel:2022wya}). The VEV of $\sigma$ breaks $B-L$ and generates masses for the right-handed neutrinos. It also gives rise to $B-L$ violating trilinear couplings between various sextet scalars which can induce neutral $n$-$\bar{n}$ oscillations through dim-9 six-fermion operators \cite{Rao:1982gt,Caswell:1982qs,Rao:1983sd}.

To derive the effective operators relevant for neutral baryon-antibaryon oscillations, we first write the most general quartic interaction terms between three sextet scalars and $\sigma$ allowed by the SM gauge symmetry and $B-L$. They are found as: 
\beqa \label{qs}
& & \eta_{ij}\, \sigma\, \Sigma^{\alpha \beta}\, {S_i}^\gamma_{\alpha \gamma}\, {S_j}^\theta_{\beta \theta}\, \label{q1} \\
& & \eta_2\, \epsilon_{\alpha \gamma \theta}\, \sigma\, \Sigma^{\alpha \beta}\, \Sigma^{\gamma \sigma}\, {\cal S}^\theta_{\beta \sigma} \label{q2}\\
& & \eta_3\, \sigma^*\, \Sigma^*_{\alpha \beta}\, \mathbb{S}^{\alpha \sigma a}_{\sigma b}\, \mathbb{S}^{\beta \rho b}_{\rho  a}\, \label{q3} \\
& & \eta_4\, \sigma^*\, \Sigma^*_{\alpha \beta}\, \overline{S}^{\alpha \sigma}_{\sigma}\, \overline{S}^{\beta \rho}_{\rho}\,. \label{q4} \eeqa
The terms in the first two lines above can arise from a quartic term $(\overline{\bf 126}_H)^4$ and mixing between $S$ and $\tilde{S}$. The third and fourth line terms can result from the gauge invariant terms $(\overline{\bf 126}_H^\dagger \overline{\bf 126}_H)^2$ and $(\overline{\bf 126}_H^\dagger {\bf 120}_H)^2$, respectively. In Eqs. (\ref{q1}-\ref{q4}), all the sextet fields are written in a mass basis. Eq. (\ref{q1}) denotes three distinct operators corresponding to $i,j = 1,2$ with $\eta_{12} = \eta_{21}$. Since $M_{S_1} < M_{S_2}$, the dominant contribution to the neutral baryon-antibaryon transition generically arises from the term corresponding to $\eta_{11}$ in Eq. (\ref{q1}). Therefore, we consider only the $S_1$ induced operators in the following.

Integrating out various colour sextet fields from the operators listed in Eqs. (\ref{q1}-\ref{q4}) and their hermitian conjugate terms and using the diquark couplings evaluated in Eqs. (\ref{sextet_126},\ref{sextet_120}), we find the following effective Lagrangian that gives rise to the baryon-antibaryon oscillation at the leading order.
\be \label{L_nn}
{\cal L}_{\rm eff}^{|\Delta B|=2} = \sum_{i=1}^3\,c_i\,{\cal O}_i\,+\,{\rm h.c.}\,, \ee
with 
\beqa \label{O_i}
{\cal O}_1 & = & \frac{1}{2}\epsilon^{\alpha \beta \gamma} \epsilon^{\sigma \rho \eta}\,\left(u^{C \dagger}_{\alpha A}\,C^{-1}\,d^{C *}_{\beta B}\right) \left(u^{C \dagger}_{\sigma C}\,C^{-1}\,d^{C *}_{\rho D}\right)
\left(d^{C \dagger}_{\gamma E}\,C^{-1}\,d^{C *}_{\eta F}\right)\,, \nonumber \\
{\cal O}_2 & = & \epsilon^{\alpha \beta \gamma} \epsilon^{\sigma \rho \eta}\,\left(d^{C \dagger}_{\alpha A}\,C^{-1}\,d^{C *}_{\sigma B}\right) \left(d^{C \dagger}_{\beta C}\,C^{-1}\,d^{C *}_{\rho D}\right)
\left(u^{C \dagger}_{\gamma E}\,C^{-1}\,u^{C *}_{\eta F}\right)\,, \nonumber \\
{\cal O}_3 & = & \frac{1}{2}\epsilon^{\alpha \beta \gamma} \epsilon^{\sigma \rho \eta}\,\left(u^{\dagger}_{\alpha A}\,C^{-1}\,d^{*}_{\beta B}\right) \left(u^{\dagger}_{\sigma C}\,C^{-1}\,d^{*}_{\rho D}\right)
\left(d^{C T}_{\gamma E}\,C^{-1}\,d^{C}_{\eta F}\right)\,.\eeqa
Here, the quark fields are written in a physical basis. The operators ${\cal O}_1$ and ${\cal O}_2$ arise from the quartic terms Eq. (\ref{q1}) and Eq. (\ref{q2}), respectively. The remaining terms, Eqs. (\ref{q3},\ref{q4}), both lead to a single operator denoted by ${\cal O}_3$.

The coefficients $c_i$ defined in Eq. (\ref{L_nn}) are determined as:
\beqa \label{c_i}
c_1 & = & \frac{2 \eta_{11} v_\sigma}{M_\Sigma^2 M_{S_1}^4}\,\left(U_{d^C}^\dagger F^* U_{d^C}^* \right)_{EF}\,\Big[\frac{4 i c_\theta^2}{15 \sqrt{15}}\, \left(U_{u^C}^\dagger F^* U_{d^C}^* \right)_{AB} \left(U_{u^C}^\dagger F^* U_{d^C}^* \right)_{CD} \Big. \nonumber \\
& + &  \frac{8 i s_\theta^2}{3 \sqrt{15}}\, \left(U_{u^C}^\dagger G^* U_{d^C}^* \right)_{AB} \left(U_{u^C}^\dagger G^* U_{d^C}^* \right)_{CD} \nonumber \\
& - & \Big. \frac{8 \sqrt{2} i c_\theta s_\theta}{15 \sqrt{3}}\, \left(U_{u^C}^\dagger F^* U_{d^C}^* \right)_{AB} \left(U_{u^C}^\dagger G^* U_{d^C}^* \right)_{CD} \Big]\,, \nonumber \\
c_2 & = & \frac{4 i}{15 \sqrt{15}}\,\frac{\eta_2 v_\sigma}{M_\Sigma^4 M_{\cal S}^2}\,\left(U_{d^C}^\dagger F^* U_{d^C}^* \right)_{AB}\,\left(U_{d^C}^\dagger F^* U_{d^C}^* \right)_{CD}\,\left(U_{u^C}^\dagger F^* U_{u^C}^* \right)_{EF}\, \nonumber \\
c_3 & = & \frac{2 v_\sigma}{M_\Sigma^2}\,\left(U_{d^C}^T F U_{d^C} \right)_{EF}\,\Big[\frac{24 i \eta_3}{15 \sqrt{15} M_{\mathbb{S}}^4}\, \left(U_{u}^\dagger F^* U_{d}^* \right)_{AB} \left(U_{u}^\dagger F^* U_{d}^* \right)_{CD} \Big. \nonumber \\
& - &\Big. \frac{8 i \eta_4}{3 \sqrt{15} M_{\overline{S}}^4}\, \left(U_{u}^\dagger G^* U_{d}^* \right)_{AB} \left(U_{u}^\dagger G^* U_{d}^* \right)_{CD} \Big]\,. \eeqa
Here, $v_\sigma = \langle \sigma \rangle$ which breaks $B-L$ by two units. The unitary matrices $U_f$ and $U_{f^C}$ with $f=u,d$ denote rotations in the flavour space and they can be explicitly computed from the corresponding quark mass matrices as described in detail in \cite{Patel:2022wya}. It can be noted that $c_1$ and $c_3$ receive contributions from both $\overline{\bf 126}_H$ and ${\bf 120}_H$.

To identify the above operators with the ones listed in \cite{Grojean:2018fus,Buchoff:2015qwa}, we rewrite various quark fields in the left- and right-chiral notations using the following relations:
\be \label{id}
\psi^T\,C^{-1}\,\chi = \overline{(\psi_L)^C}\, \chi_L\,,~~~\psi^{C\,{\dagger}}\, C^{-1}\, \chi^{C*} = \overline{(\psi_R)^C}\, \chi_R\,.\ee
Substituting the above in Eq. (\ref{O_i}), we obtain 
\beqa \label{O_i_2}
{\cal O}_1 & = & \frac{1}{2}\epsilon^{\alpha \beta \gamma} \epsilon^{\sigma \rho \eta}\,\left(\overline{(u_{R \alpha A})^C}\, d_{R \beta B}\right) \left(\overline{(u_{R \sigma C})^C}\, d_{R \rho D}\right) \left(\overline{(d_{R \gamma E})^C}\, d_{R \eta F}\right) \,, \nonumber \\
{\cal O}_2 & = & \epsilon^{\alpha \beta \gamma} \epsilon^{\sigma \rho \eta}\,\left(\overline{(d_{R\alpha A})^C}\, d_{R \sigma B}\right) \left(\overline{(d_{R \beta C})^C}\, d_{R \rho D}\right) \left(\overline{(u_{R \gamma E})^C}\, u_{R \eta F}\right)\,, \nonumber \\
{\cal O}_3^{*} & = & \frac{1}{2}\epsilon^{\alpha \beta \gamma} \epsilon^{\sigma \rho \eta}\,\left(\overline{(u_{L \alpha A})^C}\, d_{L \beta B}\right) \left(\overline{(u_{L\sigma C})^C}\, d_{L \rho D}\right) \left(\overline{(d_{R \gamma E })^C}\, d_{R \eta F}\right)\,.\eeqa
For $n$-$\bar{n}$ oscillations, one substitutes $A=B=C=D=E=F=1$ in the above expressions of ${\cal O}_i$ and $c_i$.  In this case, the operators ${\cal O}_1$ and ${\cal O}^*_3$ can straightforwardly be identified with ${\cal O}_1$ and ${\cal O}_3$ as listed in \cite{Grojean:2018fus,Buchoff:2015qwa}, respectively. Our ${\cal{O}}_2$ is proportional to ${\cal O}^1_{RRR}$ defined in \cite{Buchoff:2015qwa} which in turn is a linear combination of the operators ${\cal O}_1$ and ${\cal O}_4$ also defined in \cite{Buchoff:2015qwa}. Explicitly,
\be \label{O2_O1}
{\cal O}_2 = \frac{1}{4} {\cal O}^1_{RRR} = \frac{1}{5} \left({\cal O}_4 - 12\, {\cal \tilde{O}}_1 \right)\,.\ee
Notably, ${\cal O}_4$ has vanishing nuclear matrix element \cite{Syritsyn:2016ijx} while ${\cal \tilde{O}}_1 $ has nuclear matrix element identical to that of ${\cal O}_1$. Therefore, the operator ${\cal O}_2$ is directly related to the operator ${\cal O}_1$ in our case. This leaves only two linearly independent operators, ${\cal O}_1$ and ${\cal O}_3$, as listed in Eq. (\ref{O_i_2}).

The operators ${\cal O}_{1,3}$ need to be evolved from the scale of sextet masses, namely $\mu = M_\Phi$, down to $\mu_0 = 2$ GeV, where nuclear matrix elements are computed using the lattice calculations. The noteworthy feature about the basis in which ${\cal O}_{1,3}$ are written is that they do not mix through renormalization group evolution. The mean lifetime of the $n$-$\bar{n}$ transition can be computed in terms of the effective operators as
\beqa \label{tau_nn}
\tau_{n \bar{n}}^{-1} &=& \left|\sum_{i=1,3} \langle \bar{n}|{\cal O}_i(\mu_0)|n\rangle\, c_i(\mu_0) \right|\,,\nonumber\\
&=& \left|\sum_{i=1,3} \langle \bar{n}|{\cal O}_i(\mu_0)|n\rangle\left(\frac{\alpha_s^{(4)}(m_b)}{\alpha_s^{(4)}(\mu_0)}\right)^{\frac{3 \gamma^{(0)}_i}{50}}\left(\frac{\alpha_s^{(5)}(m_t)}{\alpha_s^{(5)}(m_b)}\right)^{\frac{3 \gamma^{(0)}_i}{46}} \left(\frac{\alpha_s^{(6)}(M_\Phi)}{\alpha_s^{(6)}(m_t)}\right)^{\frac{\gamma^{(0)}_i}{14}} c_i(M_\Phi) \right|\, \eeqa
where $\gamma^{(0)}_i$ is the leading order anomalous dimension of operator ${\cal O}_i$ and $\alpha_s^{(n_f)}$ is the strong coupling constant with $n_f$ flavours of the light quarks. We have $\gamma^{(0)}_1 = 4$ and $\gamma^{(0)}_3 = 0$ from \cite{Buchoff:2015qwa}. Using the lattice calculation results from \cite{Syritsyn:2016ijx} and parametrizing the running effects as in \cite{Grojean:2018fus}, one finally finds
\be \label{tau_nn_2}
\tau_{n \bar{n}}^{-1} = \left|0.760\, \left(\frac{\alpha_s(M_\Phi)}{\alpha_s(10^5\,{\rm GeV})} \right)^{2/7}\, \left( c_1 - \frac{12}{5}\, c_2 \right)\,+\, 1.08\, c_3^* \right|\,\Lambda_{\rm QCD}^6\,. \ee
The $n$-$\bar{n}$ oscillation time in the given model can be explicitly computed by substituting $c_{1,3}$ from Eq. (\ref{c_i}) in the above expression.

Note that $(U_{u}^\dagger G^* U_{d}^*)_{11} \neq 0$ despite of $G$ being anti-symmetric in the flavour space. This follows from the fact that $U_u \neq U_d$ in general, which in fact is necessarily required by the  realistic quark mixing. Therefore, the colour sextet scalars from ${\bf 120}_H$ can induce non-vanishing contribution to $n$-$\bar{n}$ oscillation.

\section{Perturbativity of the effective quartic couplings}
\label{sec:quartic}
\begin{figure}[t]
\centering
\subfigure{\includegraphics[width=0.30\textwidth]{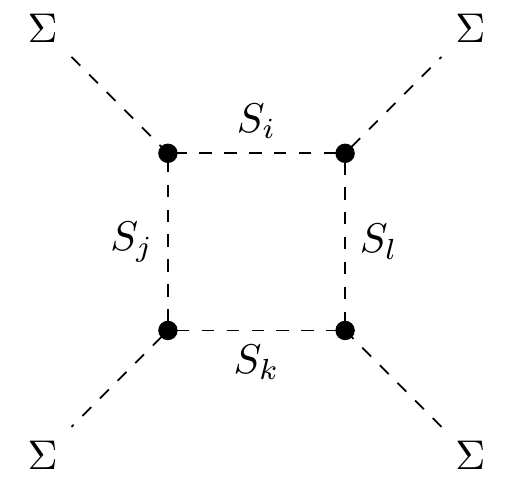}}
\caption{Diagram representing a typical correction to the quartic coupling induced by trilinear coupling between three colour sextet scalars. The vertex denoted by the bullet point is a trilinear coupling which arises from the VEV of $B-L$ charged scalar $\sigma$. }
\label{fig1}
\end{figure}

It can be seen from the expressions of $c_i$ in Eq. (\ref{c_i}) that the maximization of $n$-$\bar{n}$ transition rate would require large $v_\sigma$ and at least two colour sextet fields at the low scale. However, this possibility is known to lead to large negative effective quartic couplings for the light scalars \cite{Babu:2002uu}. It arises from the correction induced by the trilinear terms which get generated when $\sigma$ acquires VEV in Eqs. (\ref{q1}-\ref{q4}). To quantify the constraint on the masses of the underlying scalars, we consider the first term, Eq. (\ref{q1}), and compute the correction to the quartic coupling of $\Sigma$ arising from the diagram shown in Fig. \ref{fig1}.  

The effective quartic coupling, arising from a tree-level coupling $\lambda$ and the higher order corrections, can be written as:
\be \label{lambda_eff}
-i\,\lambda_{\rm eff} = -i\,\lambda + i\, \delta \lambda\,.\ee
$\delta \lambda$ computed from the diagram shown in Fig. \ref{fig1}, for the vanishing external momentum, is given by 
\be \label{dl}
 i\, \delta\lambda = 4\sum_{i,j,k,l}\, \eta_{ij} \eta_{jk}^* \eta_{kl} \eta_{li}^*\, v_\sigma^4\, \underbrace{\int \frac{d^4p}{(2 \pi)^4} \frac{1}{(p^2 - M_{S_i}^2) (p^2 - M_{S_j}^2) (p^2 - M_{S_k}^2) (p^2 - M_{S_l}^2)}}_{\equiv I_{ijkl}}\,.\ee
A straightforward computation for $i,j,k,l=1,2$ leads to
\beqa \label{dl2}
i\, \frac{\delta \lambda}{4 v_\sigma^4} &=& |\eta_{11}|^4\, I_{1111} + |\eta_{22}|^4\, I_{2222} + 4|\eta_{12}|^2 \left( |\eta_{11}|^2\, I_{1112} + |\eta_{22}|^2\,I_{2221}\right) \, \nonumber \\
& + & \left(2 |\eta_{12}|^4 + 4\, {\rm Re}[\eta_{11} \eta_{22} \eta_{12}^* \eta_{12}^*]\right)\, I_{1122}\,,\eeqa
where the different integrals are determined as \cite{Buras:2002vd}
\beqa \label{Int}
I_{iiii} &=& \frac{i}{4 \pi^2}\,\frac{1}{24 M_{S_i}^4}\,, \nonumber \\
I_{iiij} &=& \frac{i}{4 \pi^2}\,\frac{M_{S_i}^4 - M_{S_j}^4 + 2 M_{S_i}^2 M_{S_j}^2\, \log(M_{S_j}^2/M_{S_i}^2)}{8 M_{S_i}^2 (M_{S_i}^2 - M_{S_j}^2)^3}\,, \nonumber \\
I_{iijj} &=& \frac{i}{4 \pi^2}\,\frac{1}{4 (M_{S_i}^2 - M_{S_j}^2)^2}\,\left(\frac{M_{S_i}^2 + M_{S_j}^2}{M_{S_i}^2 - M_{S_j}^2}\, \log\left(\frac{M_{S_i}^2}{M_{S_j}^2}\right) - 2\right)\,.\eeqa

It can be seen that all integrals divided by  $i$ are positive. The only negative contribution to $\delta \lambda$ can come from the last term in Eq. (\ref{dl2}). It is apparent that this term cannot cancel completely the positive contributions coming from the remaining terms. Therefore, $\delta \lambda$ is always positive.
For a hierarchical $M_{S_1} \ll M_{S_2}$, one finds
\be \label{}
\lambda_{\rm eff} \simeq \lambda - \frac{|\eta_{11}|^4}{24 \pi^2}\,\frac{v_\sigma^4}{M_{S_1}^4}\,.\ee
Since both $\lambda_{\rm eff}$ and $\lambda$ are required to be positive and perturbative, the above leads to a constraint
\be \label{}
M_{S_1} \geq \frac{|\eta_{11}| v_\sigma}{(24 \pi^2)^{1/4}}\,,\ee
if $M_{\Sigma} < v_\sigma$. For $M_{\Sigma} > v_\sigma$, the above constraint does not apply as $\Sigma$ has to be integrated out first and the effective theory below $v_\sigma$ does not contain a quartic term for $\Sigma$.

It is straightforward to generalize the above discussion for the remaining sextet scalars and their interactions given in Eqs. (\ref{q1}-\ref{q4}). Generically, two or more sextet scalars coupled through $B-L$ violating vertex leads to unstable potential if they are all well below the $B-L$ breaking scale. To evade this situation, at least one of these sextets is required to be heavier than the $B-L$ breaking scale and this in turn leads to relatively suppressed $n$-$\overline{n}$ transition rate in the models with high $B-L$ breaking scale.

\section{Baryogenesis}
\label{sec:baryo}
We now point out a possibility of generating baryon asymmetry using the colour sextet scalars and their interactions predicted within this framework. The mechanism relies on the fact that the $B-L$ violating interactions present in the model can generate baryon asymmetry that avoids washout by the electroweak sphalerons \cite{Kuzmin:1987wn,Harvey:1990qw}. The sextet scalars relevant for this are $\Sigma$, $S$, $\tilde{S}$ (or $S_{1,2}$ in the physical basis) and their interactions extracted from Eqs. (\ref{sextet_all},\ref{q1}) are summarized as:
\beqa \label{Lbaryo1}
-{\cal L} &\supset &  Y^\Sigma_{AB}\,d^{C T}_{\alpha A}\,C^{-1}\,d^C_{\beta B}\, \Sigma^{\alpha \beta}\,+\, Y^{S_i}_{AB}\,\epsilon^{\alpha \beta \gamma}\, u^{C T}_{\gamma A}\,C^{-1}\,d^C_{\sigma B}\, {S_i}^\sigma_{\alpha \beta}\, \nonumber \\
& + & \eta_{ij}\, \sigma\, \Sigma^{\alpha \beta}\, {S_i}^\gamma_{\alpha \gamma}\, {S_j}^\theta_{\beta \theta}\,+\,{\rm h.c.}\,.\eeqa
The above Lagrangian contains all the necessary conditions for baryogenesis \cite{Sakharov:1967dj}. It inherently violates $P$ while $CP$ violation can arise from the phases in $\eta_{ij}$ as described in detail below. The VEV of $\sigma$ gives rise to $B-L$ violation as mentioned earlier. Departure from the thermal equilibrium is arranged through out-of-equilibrium decays of $\Sigma$ in the expanding universe as discussed below.

We assume the mass hierarchy $M_\Sigma \gg M_{S_2} > M_{S_1} \gg m_t$. The relevant processes for baryogenesis, as can be read from Eq. (\ref{Lbaryo1}), are categorized as the following.
\begin{itemize}
\item $B$ conserving: decay $\Sigma \to d^C_A\, d^C_B$ and scatterings $\Sigma\, S_i^* \to d^C_A\,\overline{u^C}_B$, $\Sigma\,u^C \to S_i\,d^C$ and $\Sigma\,\overline{d^C}_A \to S_i\,\overline{u^C}_B$,
\item $B$ violating: decay $\Sigma \to S_i^*\, S_j^*$ and scatterings $S_i^*\,S_j^* \to d^C_A\,d^C_B$, $\Sigma\, S_i \to \overline{d^C}_A\,\overline{u^C}_B$, $S_i^*\,\overline{d^C}_A \to S_j\,d^C_B$, $\Sigma\, u^C_A \to S_i^*\,\overline{d^C}_B$ and $\Sigma\,d^C_A \to S_i^*\,\overline{u^C}_B$,
\end{itemize}
along with their CP conjugate and inverse processes.

\begin{figure}[t]
\centering
\subfigure{\includegraphics[scale=1]{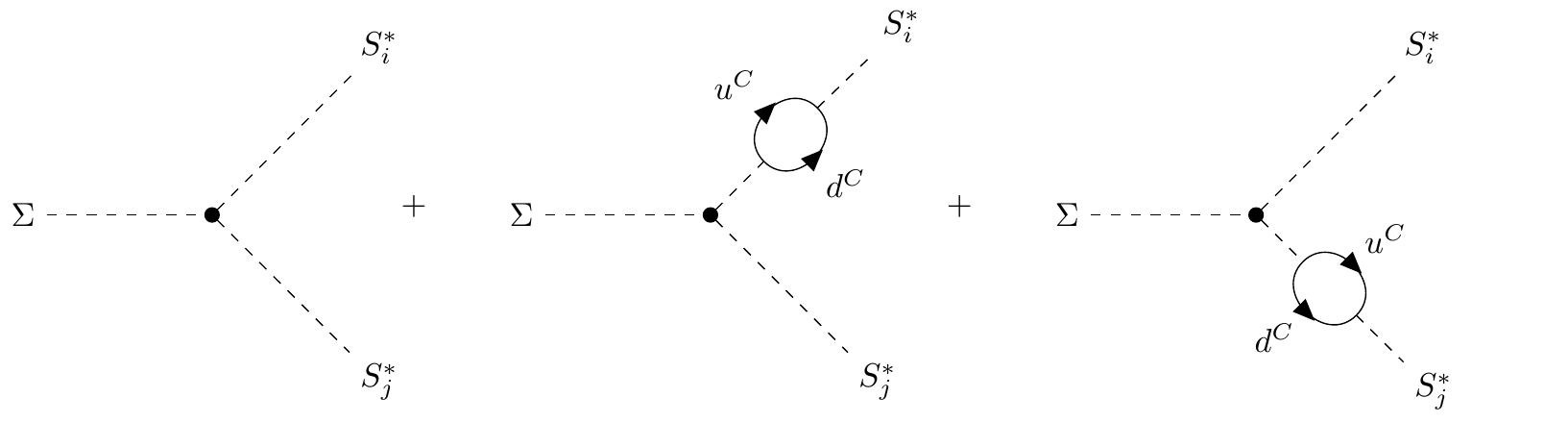}}
\caption{The tree and one-loop diagrams involving decays of the colour sextet scalar give rise to CP asymmetry. The vertex denoted by the bullet point is a trilinear coupling induced by the VEV of $B-L$ charged scalar $\sigma$. }
\label{fig2}
\end{figure}

Concentrating on the $B$ violating decay modes of $\Sigma$, the CP asymmetry in the decay $\Sigma \to S^*_i S^*_j$ is defined as
\be \label{eps1}
\epsilon_{ij} = \frac{\Gamma[\Sigma \to S^*_i S^*_j] - \Gamma[\Sigma^* \to S_i S_j]}{\Gamma_{\rm tot}[\Sigma]}\,.\ee
Nonzero $\epsilon_{ij}$ can be generated from interference between a tree and a 1-loop diagram which has an absorptive part. These diagrams are shown in Fig. \ref{fig2}.
We compute these diagrams and find the leading order CP asymmetry as 
\be \label{eps2}
\epsilon_{ij} = -\frac{1}{\pi}\, \frac{{\rm BR}[\Sigma \to S^*_i S^*_j]}{|\eta_{ij}|^2}\,\left(\sum_{k\neq i} \frac{x_{i/k}}{1-x_{i/k}}\,{\rm Im}\left[\eta^*_{ij} \eta_{jk}\,{\rm Tr}[Y^{S_i \dagger} Y^{S_j}]\right]\,+\, i \leftrightarrow j \right)\,, \ee
where $x_{i/k} = M_{S_i}^2/M_{S_k}^2$. Note that $\epsilon_{ij} = \epsilon_{ji}$ as $\eta_{ij} = \eta_{ji}$ in Eq. (\ref{Lbaryo1}). The total CP asymmetry produced in decays of $\Sigma$ is then given by
\be \label{eps_tot}
\epsilon = \epsilon_{11} + \epsilon_{22} + 2 \epsilon_{12}\,,\ee
for the present case. The asymmetry generated between the number densities of $S_i$ and $S_i^*$ due to the out-of-equilibrium decay of $\Sigma$ gets further redistributed into the SM quarks through the decays of $S_i$. This happens at the temperatures below the freeze-out of the asymmetry as $M_{S_i} \ll M_\Sigma$ and when $S_i$ leaves thermal equilibrium.

The final baryon-to-entropy ratio is obtained as
\be \label{YB}
Y_B = \frac{4}{3}\,\epsilon\, \frac{\kappa(K)}{g_*}\,, \ee
where the factor of $4/3$ is the total $B-L$ quantum number of the final states in the decay of $\Sigma$. $g_*$ is the effective number of the relativistic degrees of freedom at the time of decay. We have $g_* \simeq 125$ which include $\Sigma$, $S_{1,2}$ and the SM particles. $\kappa(K)$ is an efficiency factor which accounts for the washout of the asymmetries due to inverse decays and scattering processes listed above. $K$ is a decay parameter which is a measure of out-of-equilibrium condition and it is defined as
\be \label{K}
K = \frac{\Gamma[\Sigma \to S_i^*\,S_j^*]}{2 H(M_\Sigma)}\,,\ee
with Hubble parameter 
\be \label{H}
H(T) = 1.66\, g_*^{1/2}\,\frac{T^2}{M_P}\,.\ee

An exact value of $\kappa$ is to be obtained by numerically solving the full Boltzmann equations as outlined in \cite{Herrmann:2014fha,Fridell:2021gag}. Nevertheless, an approximate analytical solution for $\kappa$ exists which is suitable for the present setup. Assuming the initial thermal abundance for $\Sigma$, it is given by \cite{Buchmuller:2004nz}
\be \label{kappa}
\kappa(x) = \frac{2}{x\, z_B(x)} \left(1-\exp\left[-\frac{1}{2}\, x\, z_B(x)\right]\right)\,,
\ee
with
\be \label{zB}
z_B(x) = 2 + 4\, x^{0.13}\, \exp\left[-\frac{2.5}{x} \right]\,.\ee
The above solution of $\kappa$ takes into account the washout effects only by the inverse decay and it is valid for $K \le 10^3$. For $K > 10^3$, the scattering processes become more important and $\kappa$ decreases exponentially. Note that for $K \leq 1$, $\kappa(K) \to 1$ implies no dilution in the baryon asymmetry due to washout.

We aim to show that there is enough CP violation available through Eq. (\ref{eps_tot}) in the present framework such that it can account for the observed baryon to entropy ratio $Y_B^{\rm exp} = (6.10 \pm 0.04)\times 10^{-10}$ \cite{Planck:2018vyg}. For this, we first find the maximum possible value of $\epsilon$ and then evaluate the amount of damping permitted through Eq. (\ref{YB}) requiring that $Y_B \geq 6.0 \times 10^{-10}$. Assuming $\eta_{ij} = |\eta| e^{i \phi_{ij}}$ and ${\rm BR}[\Sigma \to S^*_i S^*_j] \simeq 1/4$ for all $i$ and $j$ in Eqs. (\ref{eps2},\ref{eps_tot}), we get after some straightforward algebra: 
\be \label{eps_tot_2}
\epsilon = \frac{1}{2\pi} \left(\frac{1+x_{1/2}}{1-x_{1/2}}\right)\, \left|{\rm Tr}[Y^{S_1 \dagger} Y^{S_2}]\right|\,\left(\sin(\phi_{12}-\phi_{22} - \phi_{Y}) - \sin(\phi_{12}-\phi_{11}+\phi_{Y}) \right)\,, \ee
where $\phi_{Y} = {\rm Arg}({\rm Tr}[Y^{S_1 \dagger} Y^{S_2}])$ is the phase that arise from the Yukawa couplings. For $x_{1/2} \ll 1$, maximization of $\epsilon$ leads to 
\be \label{eps_max}
\epsilon_{\rm max} \simeq \frac{1}{\pi} \left|{\rm Tr}[Y^{S_1 \dagger} Y^{S_2}]\right|\,. \ee
Substituting $Y^{S_{1,2}}$ from Eq. (\ref{YPhi}) and using the fact that ${\rm Tr}[F^\dagger G] = 0$ due to symmetric and antisymmetric properties of $F$ and $G$ respectively, we find
\be \label{eps_max2}
\epsilon_{\rm max} \simeq \frac{2}{3\pi} \left| \sin 2\theta\, {\rm Tr}\left[G^\dagger G - \frac{1}{10} F^\dagger F \right]\right|\,. \ee

\begin{figure}[t]
\centering
\subfigure{\includegraphics[width=0.47\textwidth]{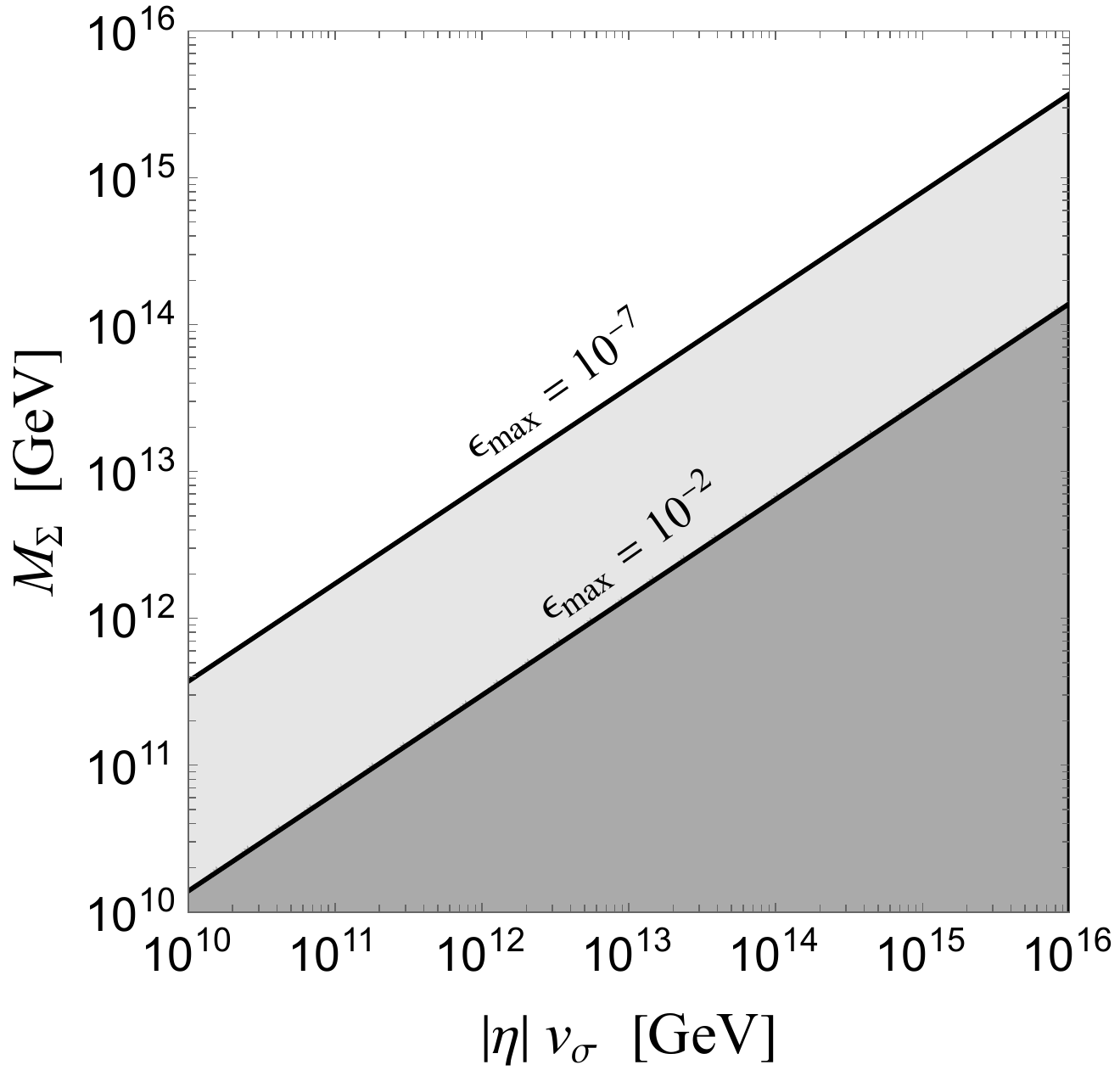}}
\caption{Values of $M_\Sigma$ and $|\eta| v_\sigma$ excluded by $Y_B > 6.0 \times 10^{-10}$ for $\epsilon_{\rm max} = 10^{-7}$ (lighter gray) and $\epsilon_{\rm max} = 10^{-2}$ (darker gray).}
\label{fig3}
\end{figure}

For $M_{S_i} \ll M_\Sigma$, one finds $\Gamma[\Sigma \to S_i^*\,S_j^*] \simeq \frac{|\eta_{ij}|^2 v_\sigma^2}{16 \pi M_\Sigma}$. Substituting this in $K$ and setting $|\eta_{ij}| = |\eta|$, we get
\be \label{K2}
K = 1.07 \times \left(\frac{|\eta| v_\sigma}{10^{15}\,{\rm GeV}}\right)^2 \times \left(\frac{10^{15}\,{\rm GeV}}{M_\Sigma} \right)^3\,.\ee
Using the above in $\kappa(K)$, the baryon to entropy ratio $Y_B$ can be computed for a given value of $\epsilon_{\rm max}$. Demanding that $Y_B \geq 6.0 \times 10^{-10}$, the allowed regions are shown in Fig. \ref{fig3}.
We observe that $M_\Sigma$ cannot be much lighter than the scale of $B-L$ breaking as it leads to washout of the baryon asymmetry through very large $K$.

The possibility of generating baryon asymmetry discussed in this section utilizes the sextets from the GUT scalars of the Yukawa sectors only and requires the presence of both $\overline{\bf 126}_H$ and ${\bf 120}_H$. Alternatively, a similar mechanism also works if only one of them is present. However, this requires an additional copy of $\Sigma$-like sextet which can emerge from ${\bf 54}$-dimensional GUT scalar \cite{Babu:2012vc}. Note that baryon asymmetry can also be generated through thermal leptogenesis as the lepton number violation and right-handed neutrinos are inherently present in the $SO(10)$ GUTs \cite{Mummidi:2021anm}.  The cases in which the latter cannot account for the observed asymmetry due to the absence of required CP violation in the lepton sector and/or suitable mass spectrum and couplings of the right-handed neutrinos, sextets governed baryogenesis discussed above can provide a viable alternative.

\section{Results}
\label{sec:results}
We now discuss constraints on the mass scales of the various colour sextet scalars from the observables quantified in the previous sections. Our emphasis is on the realistic $SO(10)$ models which are known to reproduce the observed fermion mass spectrum. It can be noted from the discussions so far that the various phenomena related to sextets involve two important parameters: (a) the Yukawa couplings with the quarks, i.e. the matrices $F$ and $G$, and (b) the $B-L$ breaking VEV $v_\sigma$. In the $SO(10)$ models with minimal choices for the scalars in the Yukawa sector, both (a) and (b) are determined from the realistic fits to the quarks and lepton masses and mixing observables \cite{Babu:1992ia,Joshipura:2011nn,Altarelli:2013aqa,Dueck:2013gca,Meloni:2014rga,Babu:2016bmy,Ohlsson:2018qpt,Mummidi:2021anm}. Using the results of the latest fit \cite{Mummidi:2021anm} performed for a non-supersymmetric $SO(10)$ model with ${\bf 10}_H$ and $\overline{\bf 126}_H$ in the Yukawa sector, one typically finds
\be \label{FG_fit}
F \sim \frac{\lambda^4}{\alpha_2} \left(\ba{ccc} \lambda^5 & \lambda^4 & \lambda^3\\
\lambda^4 & \lambda^3 & \lambda^2\\
\lambda^3 & \lambda^2 & \lambda \ea\right)\,,~~G \sim \frac{\lambda^4}{\alpha_3} \left(\ba{ccc} 0 & \lambda^4 & \lambda^3\\
- \lambda^4 & 0 & \lambda^2\\
- \lambda^3 & -\lambda^2 & 0 \ea\right)\,,\ee
where $\lambda = 0.23$ is Cabibbo angle and we have suppressed ${\cal O}(1)$ coefficients of each of the elements of $F$ and $G$. Note that the above form of $F$ is taken directly from \cite{Mummidi:2021anm} while for $G$ we assume that its non-zero elements are of similar magnitude as those of $F$ as observed in an earlier fit \cite{Joshipura:2011nn}. The parameters $\alpha_{2,3}$, with $|\alpha_{2,3}| \lesssim 1$ quantify the amount of Higgs doublet mixing as discussed in \cite{Mummidi:2021anm}.

$v_\sigma$ is determined by fitting the light neutrino masses and mixing assuming the dominance of type I seesaw mechanism and it implies
\be \label{vsgm_fit}
v_\sigma = \alpha_2\,v^\prime_S\, \simeq \alpha_2 \times 10^{15}\, {\rm GeV}\,, \ee
where $v^\prime_S$ is explicitly defined in \cite{Mummidi:2021anm} and is found in the range $10^{14}$-$10^{15}$ GeV. In addition to $v_\sigma$, $F$ and $G$, one also needs unitary matrices, $U_{f}$ and $U_{f^C}$, that diagonalize the quark mass matrices $M_f$ for $f=u,d$. They are also determined from the fits in \cite{Mummidi:2021anm} and their generic forms are given by
\be \label{U_form}
U_{u} \sim U_d \sim U_{u^C} \sim U_{d^C} \sim \left(\ba{ccc} 1 & \lambda & \lambda^3\\
\lambda & 1 & \lambda^2 \\
\lambda^3 & \lambda^2 & 1 \ea\right)\,.\ee
Again, we have suppressed the coefficient of ${\cal O}(1)$ in writing the above.

For the subsequent analysis, we consider two example values for $\alpha_{2,3}$ as the following:
\begin{itemize}
\item High-scale $B-L$ (HS): $\alpha_{2,3} = \lambda$. This implies
\be \label{FG_HS}
v_\sigma \simeq 10^{14}\,{\rm GeV}\,,~~F \simeq \left(\ba{ccc} \lambda^8 & \lambda^7 & \lambda^6\\
\lambda^7 & \lambda^6 & \lambda^5\\
\lambda^6 & \lambda^5 & \lambda^4 \ea\right)\,,~~G \simeq \left(\ba{ccc} 0 & \lambda^7 & \lambda^6\\
- \lambda^7 & 0 & \lambda^5\\
- \lambda^6 & -\lambda^5 & 0 \ea\right)\,.\ee
Since $\alpha_{2,3} = \lambda$, the lightest pair of the electroweak doublet Higgs contains a sizeable contribution from the doublets residing in $\overline{\bf 126}_H$ and ${\bf 120}_H$. $F$ and $G$ are required to be relatively small in this case. Note that $\alpha_{2,3}$ cannot be taken much greater than $\lambda$ as in that case contribution of ${\bf 10}_H$ to the fermion masses become negligible and it is known that $\overline{\bf 126}_H$ and ${\bf 120}_H$ alone cannot reproduce the observed fermion mass spectrum \cite{Joshipura:2011nn}.  
\item Intermediate-scale $B-L$ (IS): $\alpha_{2,3} = \lambda^5$. This leads to
\be \label{FG_IS}
v_\sigma \simeq 10^{11}\,{\rm GeV}\,,~~F \simeq \left(\ba{ccc} \lambda^4 & \lambda^3 & \lambda^2\\
\lambda^3 & \lambda^2 & \lambda\\
\lambda^2 & \lambda & 1 \ea\right)\,,~~G \simeq \left(\ba{ccc} 0 & \lambda^3 & \lambda^2\\
- \lambda^3 & 0 & \lambda\\
- \lambda^2 & -\lambda & 0 \ea\right)\,.\ee
In this case, $F$ and $G$ can have relatively stronger couplings with the fermions as the lightest Higgs have suppressed contributions from the doublets residing in $\overline{\bf 126}_H$ and ${\bf 120}_H$.
\end{itemize}
A low-scale $B-L$ breaking VEV, corresponding to $v_\sigma \ll 10^{11}$, would require $\alpha_{2} \ll \lambda^5$ and it makes some of the couplings in $F$ non-perturbative within this class of realistic models. Further small $\alpha_2$ with perturbative values of couplings in $F$ implies that the charged fermion masses arise dominantly from ${\bf 10}_H$ and ${\bf 120}_H$. This has been disfavoured by the fits \cite{Joshipura:2011nn}. Alternatively, the above correlation between $v_\sigma$ and the Yukawa couplings can also be understood as follows. The right-handed neutrino mass matrix is given by $M_R = v_\sigma F$ in this realistic model. The order of magnitude of the elements of $M_R$ is more or less determined by the light neutrino masses induced through the type I seesaw mechanism. This, therefore, implies smaller $F$ for the near-GUT scale $v_\sigma$ and relatively large $F$ for intermediate values of $v_\sigma$.

Before we proceed to estimate the constraints on the sextet scalars for HS and IS cases discussed above, let us outline a model-independent limit on their masses from the direct search experiments. The colour sextets can be pair-produced at the LHC from gluon fusion \cite{Chen:2008hh,Han:2009ya,Richardson:2011df}. Unlike all the observables considered in this paper, this process does not depend on the couplings with quarks and, therefore, provides a robust limit on the masses of the sextets.  Non-observation of deviation from the SM results so far implies \cite{Richardson:2011df}
\be \label{LHC}
M_{\Phi} \geq 1\,{\rm TeV}\,.\ee
The other direct search methods, such as resonant production and single top production, depend on the couplings of the sextet with quarks. They are known to provide relatively milder limits for small values of the couplings \cite{Pascual-Dias:2020hxo,Fridell:2021gag}. Therefore, we consider the above lower limit and study the other constraints in the mass range $10^3$-$10^{16}$ GeV of the sextet scalars. 

\subsection{Light $\Sigma$, $S_1$}
\begin{figure}[t]
\centering
\subfigure{\includegraphics[width=0.45\textwidth]{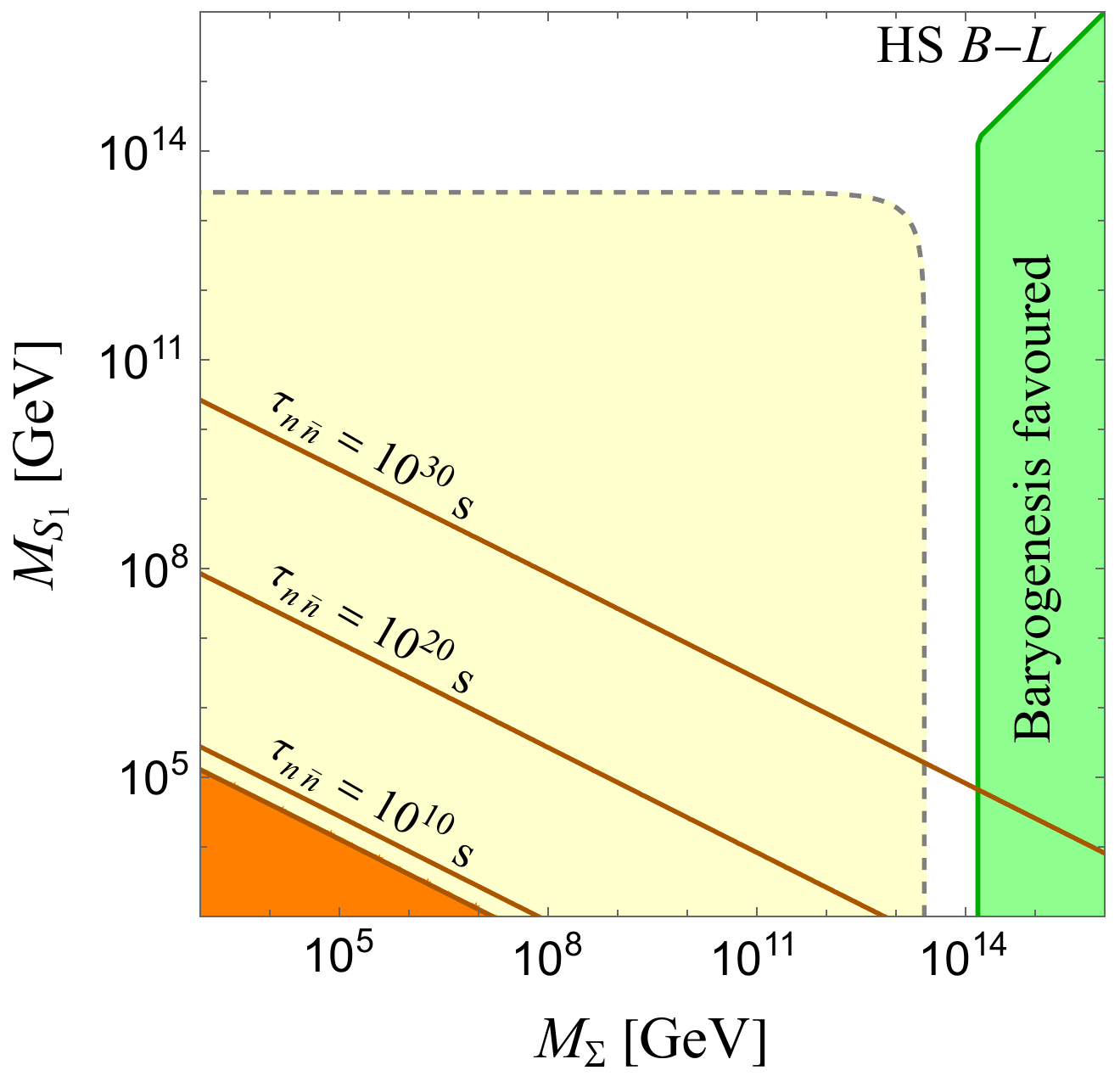}}\hspace*{0.5cm}
\subfigure{\includegraphics[width=0.45\textwidth]{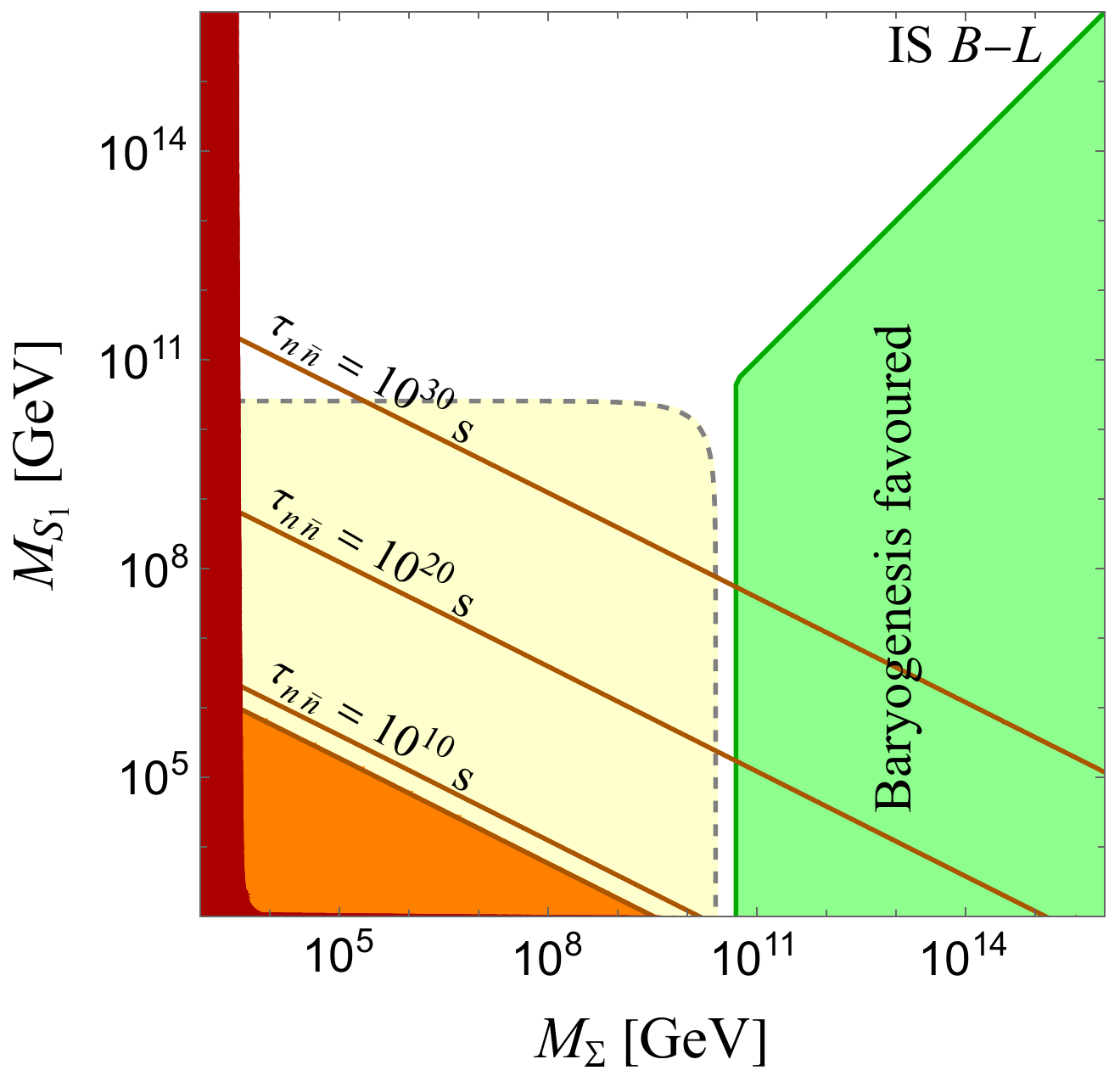}}
\caption{Constraints on the masses of $\Sigma$ and $S_1$ for the high (left panel) and intermediate (right panel) $B-L$ breaking scale. The red regions are excluded by the most dominant constraint from the neutral meson-antimeson oscillations. The region shaded in orange is excluded by the current limit $\tau_{n \bar{n}}>4.7\times 10^{8}$ seconds \cite{Super-Kamiokande:2020bov}. The diagonal red lines, from bottom to top, correspond to $\tau_{n\bar{n}} = 10^{10}$, $10^{20}$ and $10^{30}$ seconds, respectively. The yellow region bounded by the dashed contour is disfavoured by the perturbativity of the effective quartic couplings while the region shaded in green is favoured by the baryogenesis constraints.}
\label{fig4}
\end{figure}

First, we consider $1\,{\rm TeV} \le M_{\Sigma},M_{S_i} < M_{\rm GUT}$ while the remaining sextet fields stay close to the GUT scale. Using $|\eta_{11}|=1$ and $\theta=\pi/4$ in $c_1$ given in Eq. (\ref{c_i}), we compute the neutron-antineutron transition time from Eq. (\ref{tau_nn_2}) for the high- and intermediate-scale $B-L$ symmetry as described above. The relevant Yukawa couplings are evaluated using Eqs. (\ref{FG_HS},\ref{FG_IS}) and (\ref{U_form}) which give
\beqa \label{}
\left|\left(U_{d^C}^\dagger F^* U_{d^C}^* \right)_{11}\right| \sim \left|\left(U_{u^C}^\dagger F^* U_{d^C}^* \right)_{11}\right| \sim \left|\left(U_{u^C}^\dagger G^* U_{d^C}^* \right)_{11}\right| \simeq \lambda^8\,, \nonumber\\
\left|\left(U_{d^C}^\dagger F^* U_{d^C}^* \right)_{11}\right| \sim \left|\left(U_{u^C}^\dagger F^* U_{d^C}^* \right)_{11}\right| \sim \left|\left(U_{u^C}^\dagger G^* U_{d^C}^* \right)_{11}\right| \simeq \lambda^4\,,
\eeqa
for the HS and IS cases, respectively. We also compute contributions of $\Sigma$ and $S_1$ to the meson-antimeson mixing using the derived expressions, Eqs. (\ref{CK},\ref{CB},\ref{CD}), and impose the constraints from Eq. (\ref{const_C}). For the matching scale, we use $M_\Phi = (M_\Sigma + M_{S_1})/2$. The limits on $M_\Sigma$ and $M_{S_1}$ arising from meson-antimeson and neutron-antineutron oscillations are displayed in Fig. \ref{fig4}. 
We also indicate in, Fig. \ref{fig4}, a region which disfavours simultaneously light $\Sigma$ and $S_1$ due to non-perturbativity of the effective quartic coupling as discussed in section \ref{sec:quartic}.

As it can be seen from Fig. \ref{fig4}, either $\Sigma$ or $S_1$ can be substantially lighter than the GUT scale in the case of high-scale $v_\sigma$. The perturbativity of quartic coupling forbids both of them from being lighter than $10^{13}$ GeV, simultaneously. Feeble couplings with quarks allow TeV scale $\Sigma$ or $S_1$ to remain practically unconstrained from the $|\Delta F|=2$ or $|\Delta B|=2$ processes. For $v_\sigma = 10^{11}$ GeV, $M_{S_1}$ ($M_\Sigma$) can be as light as $1$ ($10$) TeV provided the other sextet is heavier then $v_\sigma$. Light $M_{S_1}$, in this case, predict relatively faster neutron-antineutron transition time in comparison to $M_\Sigma$ as can be seen from the right panel in Fig. \ref{fig4}. Overall, the constraint imposed by quartic coupling's perturbatvity almost rules out the possibility of observing $n$-$\bar{n}$ in near-future experiments for both HS and IS cases.

Sub-GUT scale $\Sigma$ ad $S_i$ can also account for the baryon asymmetry of the universe through the thermal baryogenesis as discussed in section \ref{sec:baryo}. The maximum CP asymmetry obtained using Eq. (\ref{eps_max2}) for the two cases discussed above is found to be
\be \label{}
\epsilon_{\rm max} =  \begin{cases}
      {\cal O} (10^{-7}) & \text{for HS}\\
      {\cal O} (10^{-2}) & \text{for IS}\\
    \end{cases}       \ee 
As it can be read from Fig. \ref{fig3}, the above implies
\be \label{}
M_\Sigma \gtrsim  \begin{cases}
      2 \times 10^{14}\,{\rm GeV} & \text{for HS}\\
      5 \times 10^{10}\,{\rm GeV} & \text{for IS}\\
    \end{cases}       \ee 
such that  $Y_B \geq 6.0 \times 10^{-10}$. This region favoured by sextet-generated baryogenesis is also shown in Fig. \ref{fig4}. For the very light $S_1$ and $v_\sigma  \simeq 10^{11}$ GeV, this region can be probed through improved measurements of $n$-$\bar{n}$ oscillations.

\subsection{Light $\Sigma$, ${\cal S}$}
\begin{figure}[t]
\centering
\subfigure{\includegraphics[width=0.45\textwidth]{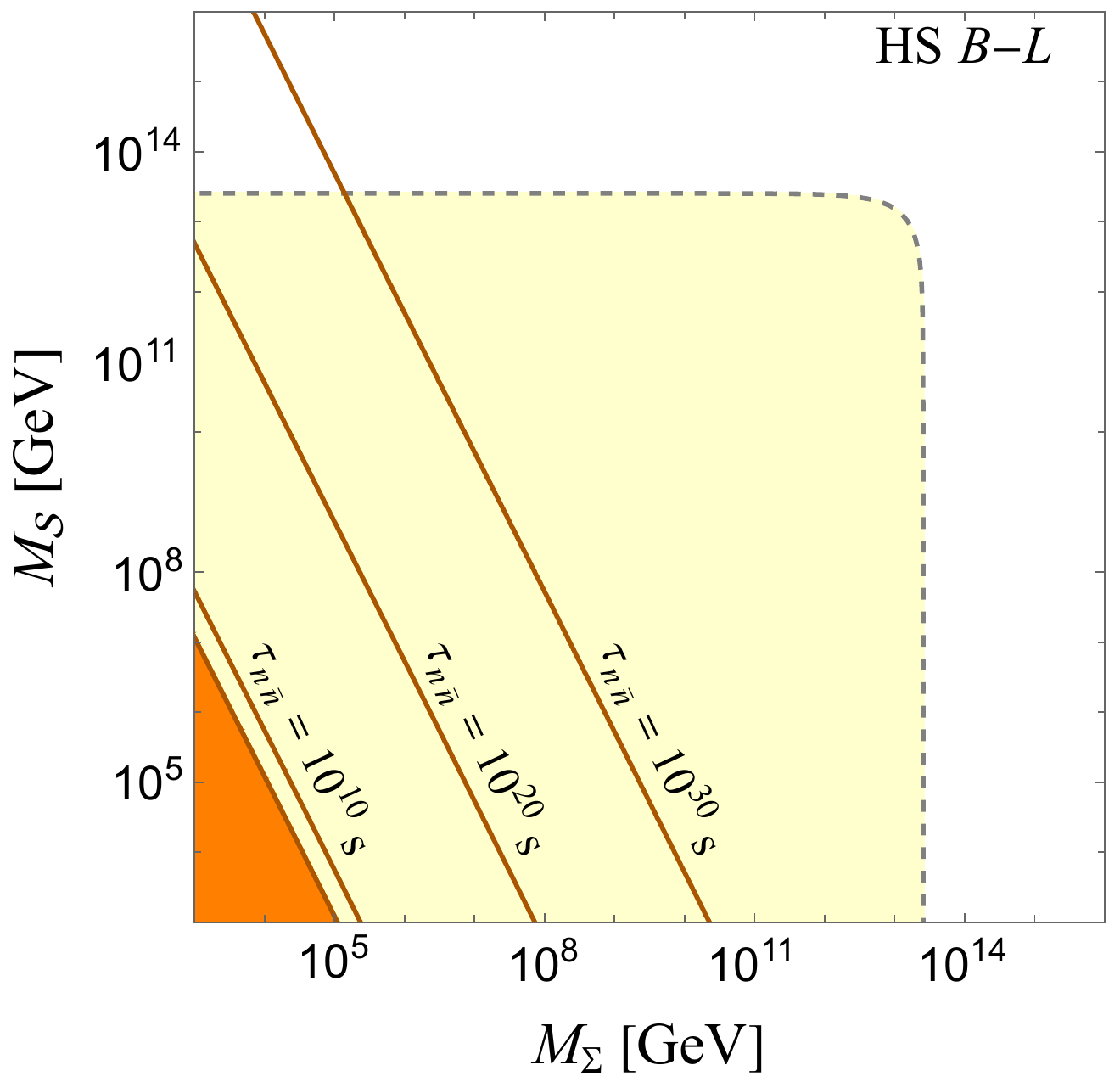}}\hspace*{0.5cm}
\subfigure{\includegraphics[width=0.45\textwidth]{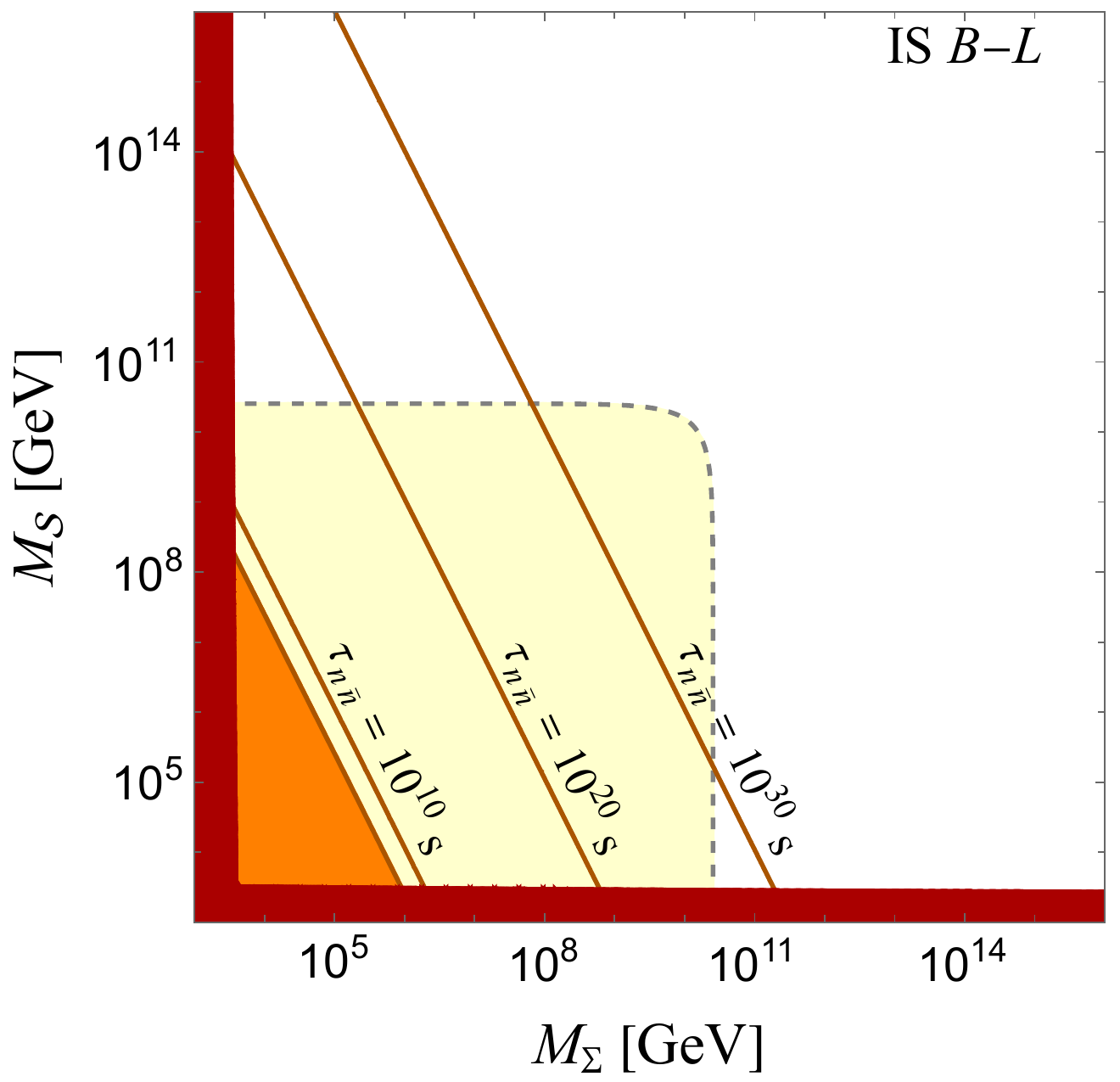}}
\caption{Constraints on the masses of $\Sigma$ and ${\cal S}$ for the high (left panel) and intermediate (right panel) $B-L$ breaking scale. The other details are the same as the caption of Fig. \ref{fig4}.}
\label{fig5}
\end{figure}

Assuming $|\eta_2| = 1$ in Eq. (\ref{q2}), we now assess the constraints on light $\Sigma$ and ${\cal S}$ assuming the remaining sextets at the GUT scale. Unlike $S_{1,2}$, ${\cal S}$ couples to the only up-type quarks and mediates \dd oscillations at the tree level. This puts severe constraints on the strongly coupled TeV scale ${\cal S}$. The constraints derived from various considerations are displayed in Fig. \ref{fig5}.
It can be seen that the limits from the meson-antimeson oscillations and perturbativity of the effective quartic coupling imply no observable $n$-$\bar{n}$ transition rate in the near-future experiments in the realistic renormalizable $SO(10)$ based models.

\subsection{Light $\Sigma$, $\mathbb{S}$}
\begin{figure}[t]
\centering
\subfigure{\includegraphics[width=0.45\textwidth]{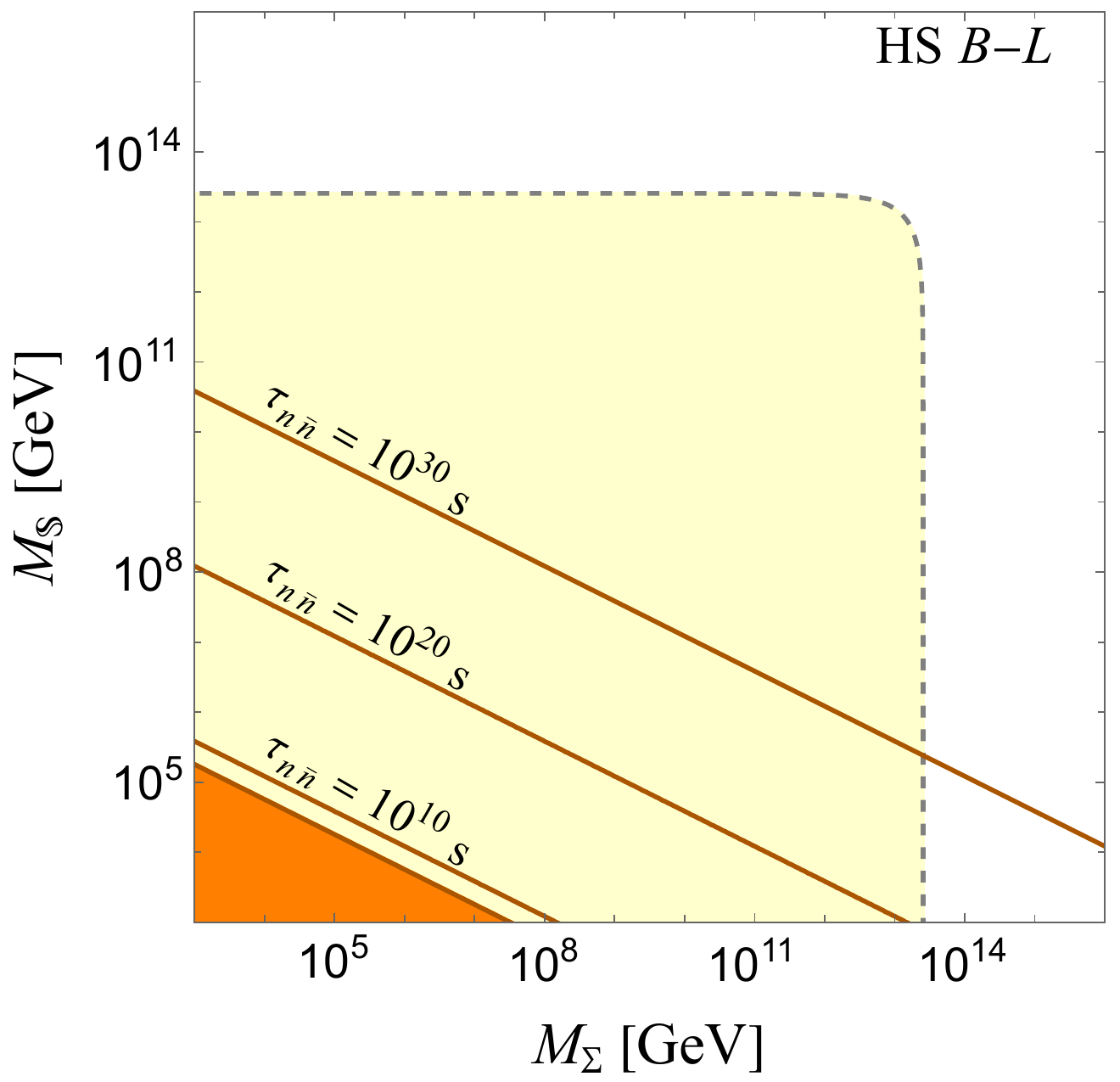}}\hspace*{0.5cm}
\subfigure{\includegraphics[width=0.45\textwidth]{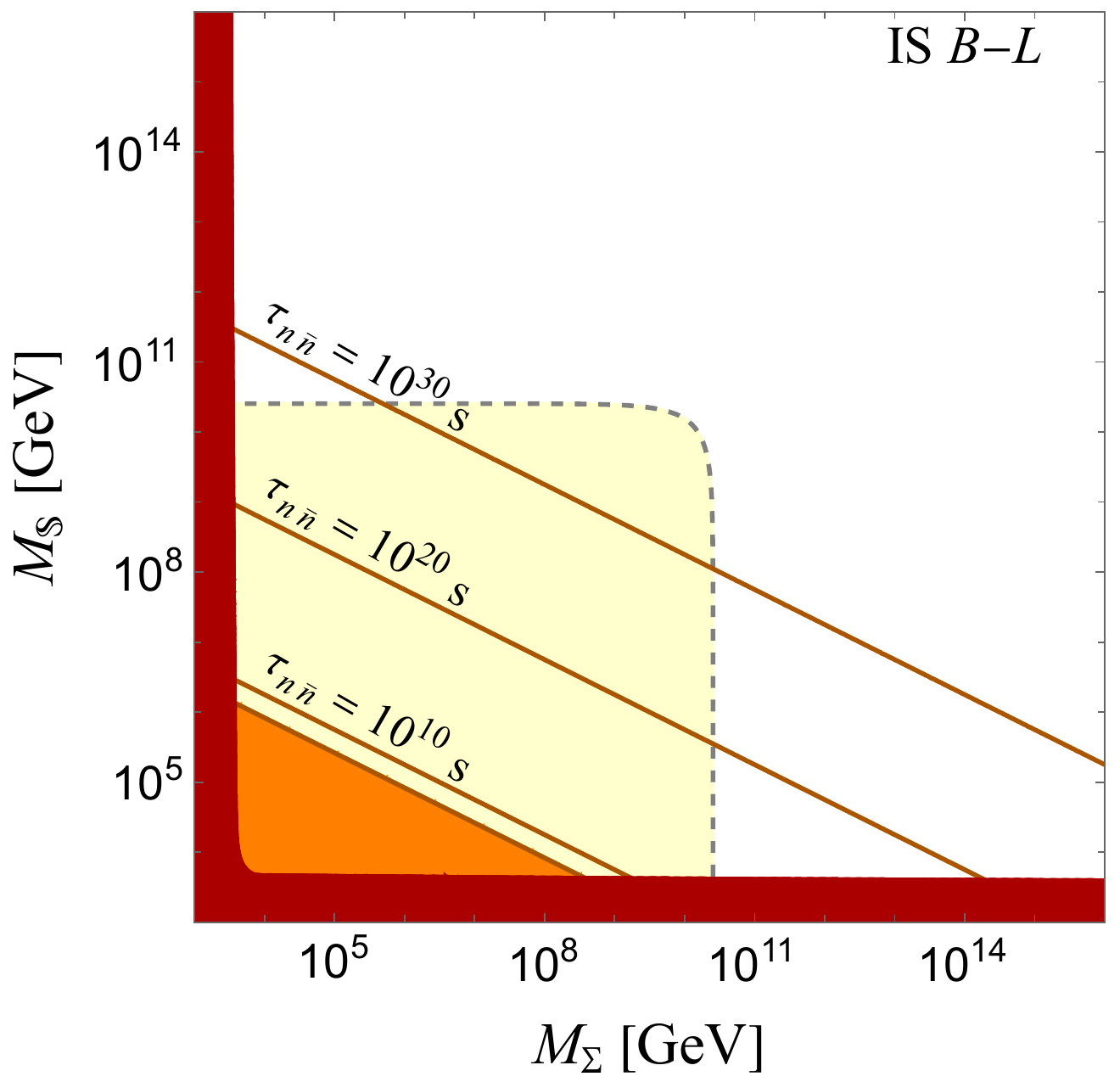}}
\caption{Constraints on the masses of $\Sigma$ and $\mathbb{S}$ for the high (left panel) and intermediate (right panel) $B-L$ breaking scale. The other details are the same as the caption of Fig. \ref{fig4}.}
\label{fig6}
\end{figure}

Next, we consider light $\Sigma$ and $\mathbb{S}$ with the remaining sextets decoupled at the GUT scale. The results are shown in Fig. \ref{fig6}. We set $|\eta_3|  =1$ for this analysis and consider two cases for the values of $F$ and $v_\sigma$ as discussed earlier. Unlike $S_i$, the electroweak triplet $\mathbb{S}$ mediate quark flavour violating interactions at tree-level in the both up- and down-type quark sector. This results in a relatively large upper bound on $M_{\mathbb{S}}$ in the case of the strong Yukawa coupling. The constraints from $n$-$\bar{n}$ oscillations and perturbativity of the effective quartic coupling are similar to the ones obtained in the case of light $\Sigma$ and $S_1$. Sub-GUT scale $\Sigma$ and $\mathbb{S}$ alone cannot generate baryon asymmetry and an additional copy of $\Sigma$-like scalar would be required if viable baryogenesis is to be realized in this case as discussed in section \ref{sec:baryo}.

\subsection{Light $\Sigma$, $\overline{S}$}
\begin{figure}[t]
\centering
\subfigure{\includegraphics[width=0.45\textwidth]{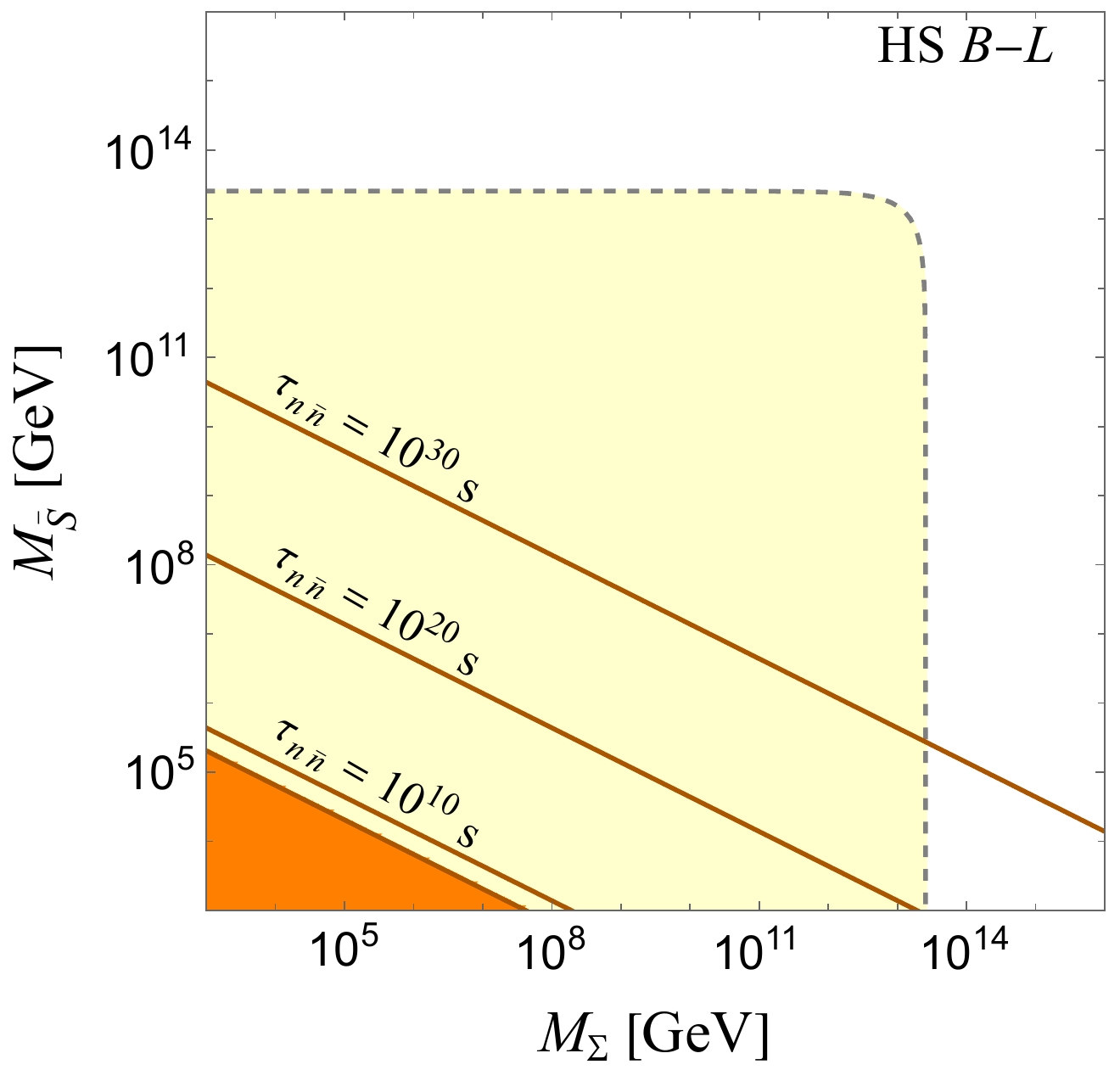}}\hspace*{0.5cm}
\subfigure{\includegraphics[width=0.45\textwidth]{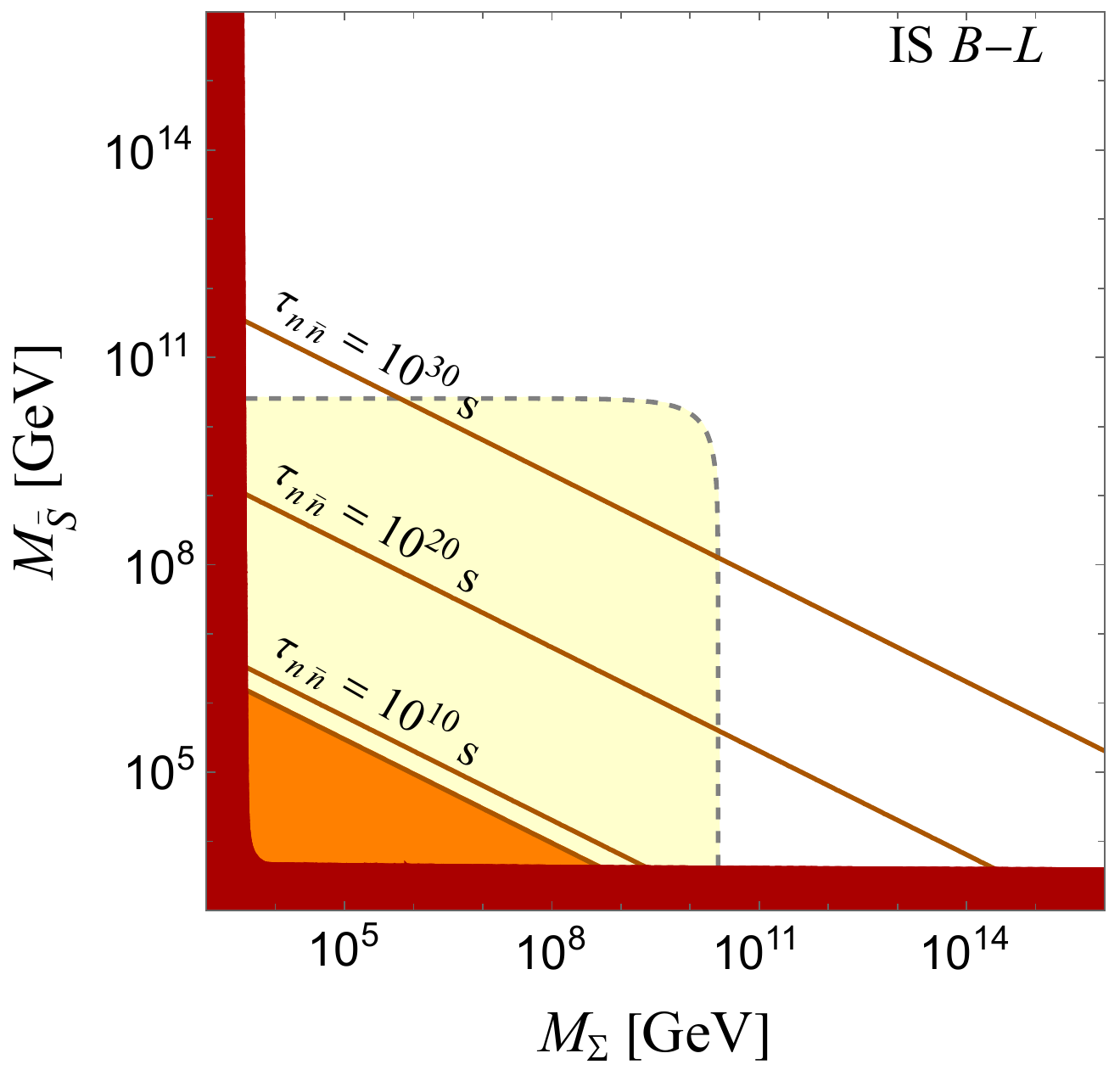}}
\caption{Constraints on the masses of $\Sigma$ and $\overline{S}$ for the high (left panel) and intermediate (right panel) $B-L$ breaking scale. The other details are the same as the caption of Fig. \ref{fig4}.}
\label{fig7}
\end{figure}

Finally, we consider a case for $1\,{\rm TeV} \le M_{\Sigma},M_{\overline{S}} < M_{\rm GUT}$ and the GUT scale masses for the remaining sextets. Setting $|\eta_4| = 1$, the obtained results are shown in Fig. \ref{fig7}. 
$\overline{S}$ originates solely from ${\bf 120}_H$ and it has flavour anti-symmetric couplings with left-chiral up- and down-type quarks. As it can be seen from $Y^{\overline{S}}$ in Eq. (\ref{YPhi}), its diagonal couplings vanish for $U_u = U_d$. The latter, however, is not supported by the non-trivial quark mixings. Hence, the effective diagonal couplings are generated by the quark mixing leading to a non-vanishing  rate for $n$-$\bar{n}$ oscillations at the leading order. Like $S_{1,2}$, $\overline{S}$ also contributes to the $|\Delta F|=2$ processes at 1-loop level. However, these processes put stronger constraints on the light $\overline{S}$ in comparison to $S_{1,2}$ as can be seen from Figs. \ref{fig4} and \ref{fig7}. This is attributed to the relatively large Yukawa coupling of $\overline{S}$ due to different Clebsch-Gordan factors. Overall, the constraints on the masses of sub-GUT scale $\mathbb{S}$ and $\overline{S}$ are similar.

\section{Conclusions}
\label{sec:concl}
Scalars which transform as two index symmetric representations of the $SU(3)_C$ of the SM gauge symmetry have been considered actively for their various phenomenological applications in bottom-up approaches. We evaluate the possibility of these sextets originating from the realistic renormalizable models of gauge and quark-lepton unification based on the $SO(10)$ symmetry. Five distinct colour sextets are naturally accommodated in this class of models: $\Sigma \sim (6,1,-\frac{2}{3})$, $S \sim (6,1,\frac{1}{3})$, $\overline{S}\sim(\overline{6},1,-\frac{1}{3})$, ${\cal S}\sim(6,1,\frac{4}{3})$ and $\mathbb{S}\sim(\overline{6},3,-\frac{1}{3})$. Deriving their couplings with the quarks, we compute effective operators contributing to the electrically neutral meson-antimeson and baryon-antibaryon oscillations. The latter arises from $B-L$ breaking induced by VEV, $v_\sigma$, of an SM singlet field $\sigma$. We also evaluate the effective quartic coupling of the sextet scalars which is prone to receive large contributions from $B-L$ breaking effects.

The noteworthy points from our present study are the following.
\begin{itemize}
\item Four pairs of the sextets, i.e. $\Sigma$-$S$, $\Sigma$-$\mathcal{S}$, $\Sigma$-$\mathbb{S}$ and $\Sigma$-$\overline{S}$, can give rise to the lifetime of neutron-antineutron transitions observable in near-future experiments provided that both the sextets in a pair are lighter than $10^5$-$10^8$ GeV. This possibility is almost entirely excluded by the perturbativity of the quartic couplings in the models with $v_\sigma \geq 10^{11}$ GeV.
\item Observable $n$-$\bar{n}$ oscillation along with perturbative effective quartic couplings can be achieved if $v_\sigma < 10^8$ GeV and couplings of the sextets with quarks are of ${\cal O}(1)$. However, this generically leads to relatively light right-handed neutrino masses inconsistent with the type I seesaw mechanism in the realistic $SO(10)$ models.
\item  For $M_\Sigma > 10^{11}$ GeV, $S$ can be as light as of ${\cal O}({\rm TeV})$ while the masses of $\mathbb{S}$, $\overline{S}$ and ${\cal S}$ can be $\gtrsim 10$ TeV. Similarly, for $S$,  $\mathbb{S}$, $\overline{S}$ and ${\cal S}$ heavier than $10^{11}$ GeV, $M_\Sigma$ can be of ${\cal O}(10)$ TeV or heavier. The lower limits on the masses of these light sextets come almost entirely from meson-antimeson oscillations and/or from direct searches.
\item $SO(10)$ models with both $\overline{\bf 126}_H$ and ${\bf 120}_H$ in the Yukawa sector leads to the existence of a pair of sextets, $S_{1,2}$, with quantum numbers identical to that of $S$. $\Sigma$ heavier than $v_\sigma$ and $S_{1,2} \ll M_{\Sigma}$, in this case, provides a novel and viable possibility of generating baryon asymmetry of the universe.
\end{itemize}

Many of the above observations follow from the fact that the couplings of sextets with quarks and the $B-L$ scale are strongly correlated in the renormalizable class of $SO(10)$ GUTs and they cannot take arbitrary values as typically assumed in the bottom-up approaches. On the other hand, a positive signal of $n$-$\bar{n}$ oscillation in near-future experiments will rule out this simplest and predictive framework of grand unification. This study provides an interesting example of how a well-defined model in the ultraviolet can lead to a restrictive class of new physics at low energies making the former a falsifiable theory.

 In the present work, our aim has been to study the phenomenological constraints on the colour sextet scalars which have direct couplings with the quarks governed by the $SO(10)$ grand unification. We do not discuss the constraints on the scalar mass spectra which may arise from the full scalar potential and requirement of consistent symmetry breaking. A careful treatment of this requires the specification of a full model beyond the Yukawa sector and it is a highly model-dependent exercise. Nevertheless, availing the freedom to choose a suitable set of GUT scalars and with an appropriate choice and tuning of parameters in the scalar potential, it is expected that the desired scalar mass spectra can be realised in concrete models.

\section*{Acknowledgements}
We acknowledge fruitful discussions with Namit Mahajan. SKS also thanks Dayanand Mishra and Gurucharan Mohanta for useful discussions. This work is partially supported under the  Mathematical Research Impact Centric Support (MATRICS) project (MTR/2021/000049) funded by the Science \& Engineering Research Board (SERB), Department of Science and Technology (DST), Government of India. 

\bibliography{references}

\end{document}